\documentclass[]{IEEEtran}
\usepackage{graphicx}
\usepackage{subfigure}
\usepackage{amsmath}
\usepackage{amsthm}
\usepackage{amssymb}
\usepackage{mathrsfs}
\usepackage{slashbox}
\usepackage{float}
\usepackage[vlined,ruled,linesnumbered]{algorithm2e}
\usepackage{xcolor}
\usepackage{verbatim}
\usepackage{booktabs}
\usepackage{multirow}

\newtheorem{thm}{Theorem}%[section]

\newtheorem{definition}{Definition}

\ifCLASSOPTIONcompsoc
  \usepackage[nocompress]{cite}
\else
  \usepackage{cite}
\fi

\ifCLASSINFOpdf

\else

\fi

\begin{document}

\title{More Than Privacy: Applying Differential Privacy in Key Areas of Artificial Intelligence}

\author{Tianqing Zhu, Dayong Ye, Wei Wang, Wanlei Zhou and Philip S. Yu*% <-this % stops a space
\thanks{Philip S. Yu is the corresponding author. Tianqing Zhu is with the School of Computer Science, China University of Geosciences, Wuhan, China; D. Ye, W. Wang and W. Zhou are with the Centre for Cyber Security and Privacy and the School of Computer Science, University of Technology, Sydney, Australia. Philip S. Yu is with the Department of Computer Science, University of Illinois at Chicago, USA. Email: Tianqing.e.zhu@gmail.com; \{Dayong.Ye, Wei.Wang-3, Wanlei.Zhou\}@uts.edu.au, psyu@cs.uic.edu}}% <-this % stops a space

%\markboth{IEEE Transactions on Knowledge and Data Engineering}%
%{Zhu \MakeLowercase{\textit{et al.}}: }

\IEEEtitleabstractindextext{%
\begin{abstract}
Artificial Intelligence (AI) has attracted a great deal of attention in recent years.
However, alongside all its advancements, problems have also emerged, such as privacy violations, security issues and model fairness.
Differential privacy, as a promising mathematical model, has several attractive properties that can help solve these problems, making it quite a valuable tool.
For this reason, differential privacy has been broadly applied in AI
but to date, no study has documented which differential privacy mechanisms can or have been leveraged to overcome its issues or the properties that make this possible.
In this paper, we show that differential privacy can do more than just privacy preservation.
It can also be used to improve security, stabilize learning, build fair models, and impose composition
in selected areas of AI.
With a focus on regular machine learning, distributed machine learning, deep learning, and multi-agent systems,
the purpose of this article is to deliver a new view on many possibilities for improving AI performance with
differential privacy techniques.
%Therefore, this paper can serve as a guide and a starting point for researchers
%who intend to conduct research on differential privacy in AI.
\end{abstract}

}

\maketitle

\IEEEdisplaynontitleabstractindextext

\IEEEpeerreviewmaketitle
\vspace{10mm}
\IEEEraisesectionheading{\section{Introduction}\label{sec:introduction}}
\IEEEPARstart{A}{rtificial} Intelligence (AI) is one of the most prevalent topics of research today across almost every scientific field.
For example, multi-agent systems can be applied to distributed control systems~\cite{GE20181684}, while distributed machine learning
has been adopted by Google for mobile users~\cite{Geyer17}.
However, as AI becomes more and more reliant on data,
several new problems have emerged, such as privacy violations, security issues, model instability, model fairness and communication overheads.
As just a few of the tactics used to derail AI,
adversarial samples can fool machine learning models,
leading to incorrect results.
Multi-agent systems may receive false information from malicious agents.
%Take the autonomous vehicle as an example.
%The essential components include machine learning models and sensor communication systems.
%In addition to the instability and unfair of the models, which would have great impact on the objective recognition or decision making,
%false information that from malicious sensors would increase the dangers of the self-driving cars;
%The privacy leakages from training data or models would disclose the drivers personal information;
%and the communication overhead would increase the processing time during the driving.
%There are lots of accidental reported in the past few years.
%Self-driving Uber car that hit and killed woman did not recognize that pedestrians jaywalk\footnote{https://www.nbcnews.com/tech/tech-news/self-driving-uber-car-hit-killed-woman-did-not-recognize-n1079281}.
%And a crash caused by Tesla autopilot system led to the death of the driver as the autonomous drving model failed to recognize the white truck againt the bright sky\footnote{https://www.cnbc.com/2019/09/03/tesla-autopilot-engaged-before-2018-california-crash-ntsb-says.html}.
%These examples show that AI security and privacy are highly related to the human security and safety.
As a result, many researchers have been exploring new and existing security and privacy tools to tackle these new emerging problems.
Differential privacy is one of these tools.

\begin{figure}[ht]
\centering
	\includegraphics[scale=0.7]{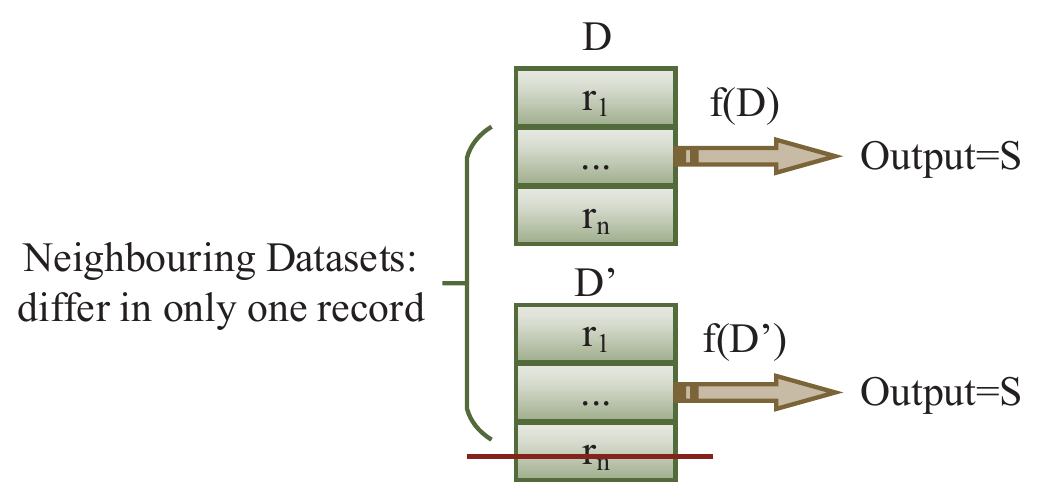}
	\caption{Differential privacy}
	\label{fig-dp}
\end{figure}

Differential privacy is a prevalent privacy preservation model which guarantees whether an individual's information is included in a dataset has little impact on the aggregate output.
Fig.~\ref{fig-dp} illustrates a basic differential privacy framework using the following example.
Consider two datasets that are almost identical but differ in only one record and that,
access to the datasets is provided via a query function $f$.
If we can find a mechanism that can query both datasets and obtain the same outputs,
we can claim that differential privacy is satisfied.
In that scenario,
an adversary cannot associate the query outputs with either of the two neighbouring datasets,
so the one different record is safe.
Hence,
the differential privacy guarantees that, even if
an adversary knows all the other records in a dataset except for one unknown individual,
they still cannot infer the information of that unknown record.

Interest in differential privacy mechanisms not only ranges from
the privacy community
to the AI community,
it has also attracted the attention of many private companies, such as Apple\footnote{https://www.apple.com/au/privacy/approach-to-privacy/
}, Uber\footnote{https://www.usenix.org/node/208168} and Google~\cite{RAPPOR14Google}.

The key idea of differential privacy is to introduce calibrated randomization to the aggregate output.
When Dwork et al.~\cite{Dwork14Science} showed that applying differential privacy mechanisms
to test data in machine learning could prevent over-fitting of learning algorithms,
it launched a new direction beyond simple privacy preservation to one that solves emerging problems in AI~\cite{Zhu20Philip}.
We use two examples to illustrate how those new properties can be applied.

\subsection{Examples}

\begin{figure}[htpb]
\centering
\includegraphics[scale=0.7]{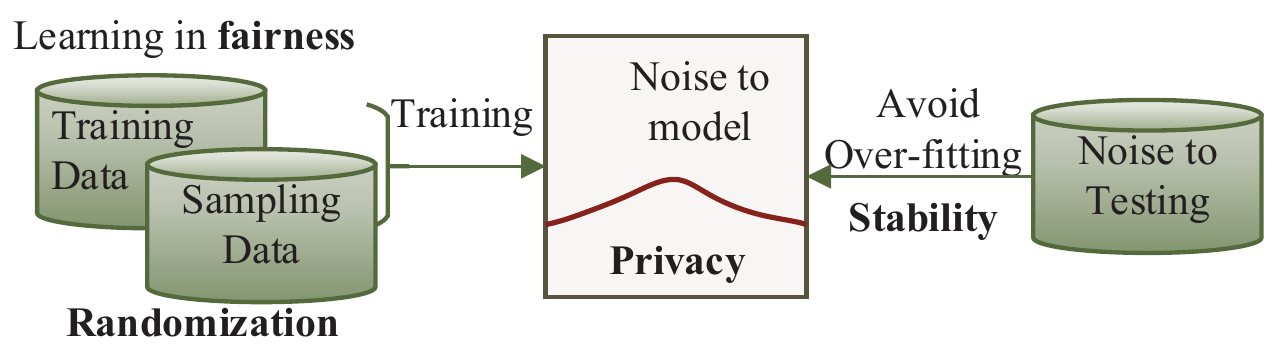}
\caption{Learning example}
\label{Fig-examplelearning}
\end{figure}

The first example pertains to machine learning.
As shown in Fig.~\ref{Fig-examplelearning},
machine learning suffers from several problems, including privacy violations, over-fitting
and unfair models.
Recent research has shown that differential privacy mechanisms have the potential to tackle those problems. First, to maintain fairness in a model, the training data can be re-sampled from the data universe using a differential privacy mechanism~\cite{Alexandra18}.
Second, to preserve privacy, noise derived from a differential privacy mechanism can be added to the learning model~\cite{Chaudhuri20111069}.
Finally,
calibrated noise can be applied to generate fresh testing data to increase stability and avoid over-fitting of the learning algorithm~\cite{Dwork14Science}.
These successful applications of differential privacy show that learning problems can be solved by taking advantage of several properties of differential privacy,
such as randomization,
privacy preservation capability, and algorithm stability.

\begin{figure}[htpb]
\centering
\includegraphics[scale=0.7]{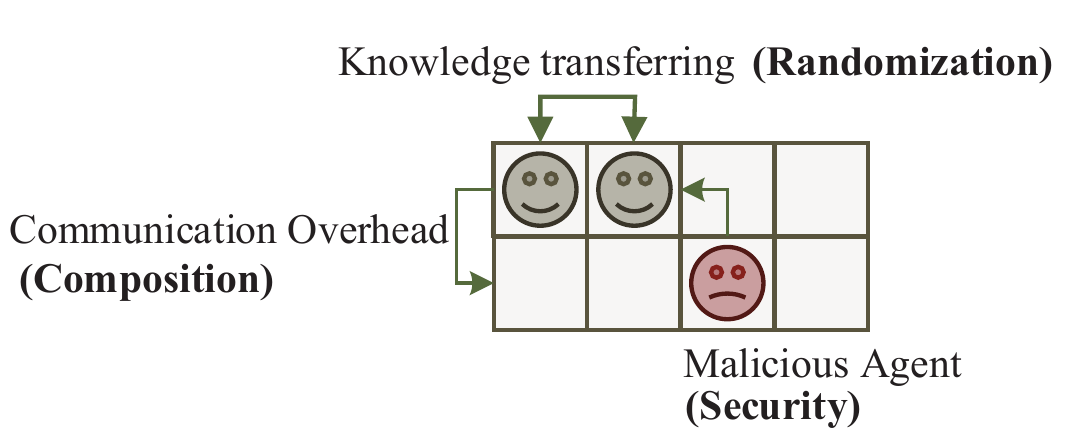}
\caption{Multi-agent example}
\label{Fig-multiagent}
\end{figure}

The second example comes from the realm of multi-agent systems, one of the traditional disciplines in AI.
A multi-agent system is a computerized system composed of multiple interacting intelligent agents,
such as sweeping robots as shown in Fig.~\ref{Fig-multiagent}.
The faces are agents and the grid denotes the moving environment of all agents.
An agent can make decisions over its direction of movement and can share that knowledge with other agents to help them make their decisions. The goal is for the robots to sweep all grids. Several problems exist in this multi-agent system. First, as each agent observes a different environment, it is difficult to share their knowledge. The randomizing mechanism in differential privacy can help to transfer the knowledge between agents. Second, communications between agents should be restricted to limit power consumption. Here, the privacy budget in differential privacy can help the system control to overall communications~\cite{Ye19}.
Third, when a malicious agent is present, like the agent in the red face, they may provide false knowledge. Differential privacy mechanisms can help improve the security level of communications by diminishing the impact of that agent.
%In other words, even if a malicious agent tries to mislead the system, differential privacy can guarantee that a malicious agent will have very little impact on the entire system.

Both these examples show how current research is applying differential privacy mechanisms to AI and how randomization can bring several new properties to AI.

\subsection{AI areas}

\begin{figure}[ht]
\centering
   \caption{AI areas in the view of acting humanly}
    \label{fig-aiarea1}
    \includegraphics[scale=0.38]{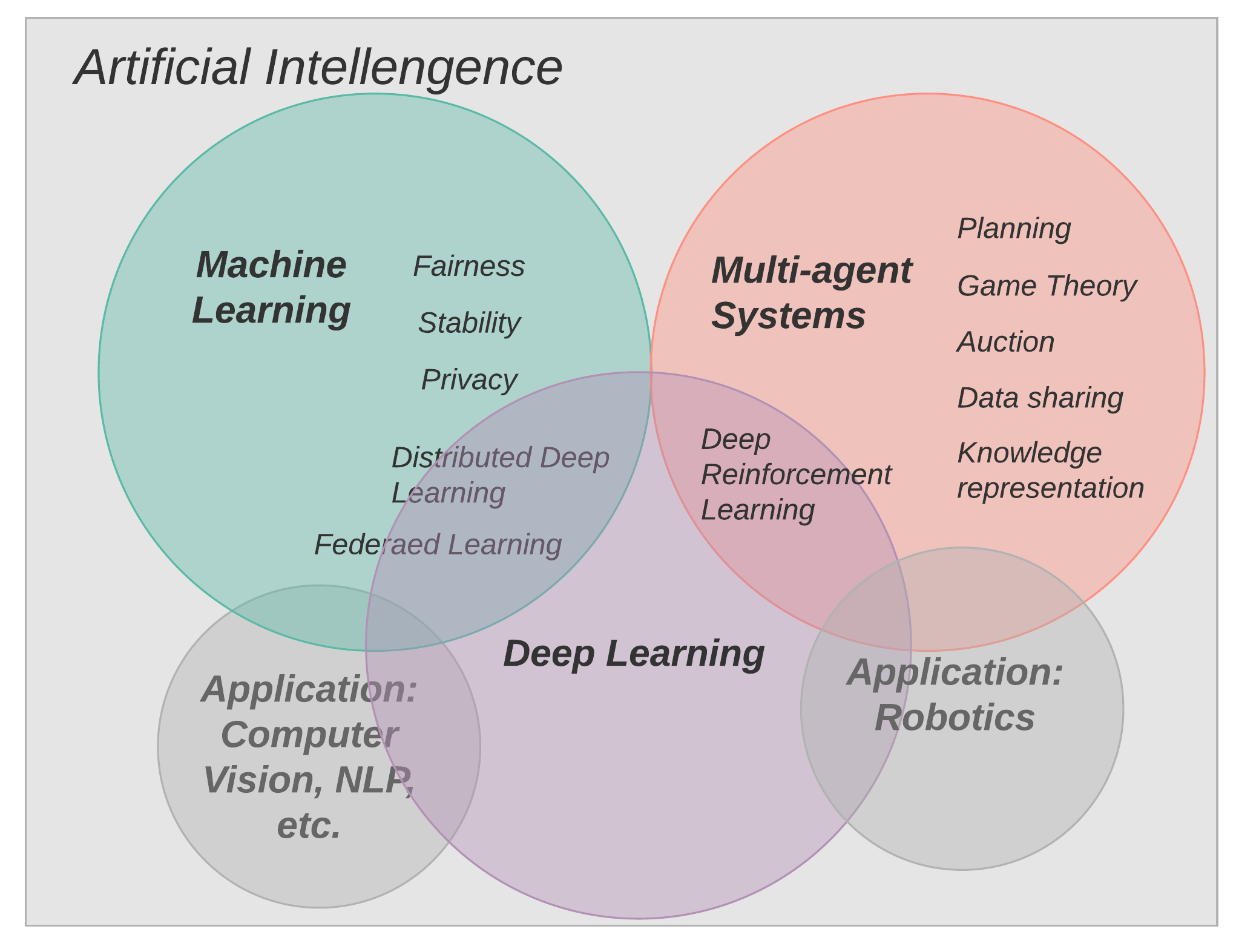}

\end{figure}

In AI, there are no strict area disciplines. Researchers and industries have built a birds-eye view of AI in all its diversity. Take the perspective of the Turing Test, for example. When programming a computer that needs to act like a human, the computer must have the following capabilities~\cite{russell2002}: natural language processing so it can successfully communicate with a human; knowledge representation to store what it knows or hears; automated reasoning to use the stored information to answer questions and to draw new conclusions; machine learning to adapt to new circumstances and to detect and extrapolate patterns; computer vision to perceive objects; and robotics to manipulate objects.

Based on this birds-eye view, we roughly categorize three major technical fields in AI: machine learning, deep learning and multi-agent system. Knowledge learning and automated reasoning can be processed by a multi-agent system; the other functions can be accomplished through machine learning and/or deep learning. In the view of the application, the AI area includes robotics, computer vision, natural language processing (NLP), etc.--see Fig~\ref{fig-aiarea1}.

Here, we note that although deep learning was originally a series of machine learning algorithms implemented in a neural network architecture, it has rapidly developed into a field of study in its own right with a huge number of novel perspectives and technologies, such as GANs~\cite{Goodfellow14}, ResNets~\cite{15RseNet}, etc. Therefore, we place deep learning in its own category.

The purpose of this paper is to document how the differential privacy mechanism can solve those new emerging problems in the technical fields: machine learning, deep learning and multi-agent systems. Applications such as robotics, NLP and computer vision have taken advantage of technologies such machine learning, deep learning and multi-agent system, so we have no reviewed these applications in dedicated sections.

\subsection{Differential privacy in AI areas}
Calibrated randomization benefits some AI algorithms. What follows is a summary of several properties derived from randomization.

\begin{itemize}
  \item \textbf{Preserving privacy}.
  This is the original purpose of differential privacy.
  By hiding an individual in the aggregate information,
  differential privacy can preserve the privacy of participants in a dataset.

  \item \textbf{Stability}.
Differential privacy mechanisms ensure that the probability
of any outcome from a learning algorithm is
unchanged by modifying any individual record in the training data.
This property
establishes connections between a
learning algorithm and its ability to be generalized.

\item \textbf{Security}.
Security relates to malicious participants in a system. Differential privacy mechanisms can reduce the impact of malicious participants in AI tasks. This property can guarantee security in AI systems.

  \item \textbf{Fairness}.
In machine learning, a given algorithm is said to be fair, or to have fairness, if its results are independent of sensitive attributes, like race and gender. Differential privacy can help to maintain fairness in a learning model by re-sampling the training data from the universe.

  \item \textbf{Composition}.
Differential privacy mechanisms can guarantee that any step that satisfies differential privacy can construct a new algorithm that also satisfies differential privacy. This property is referred to as composition and is controlled by the privacy budget. In AI, composition can be used to control the number of steps, communication loads, etc.
\end{itemize}

\begin{table*}[]
\caption{Properties of differential privacy in artificial intelligence}\label{table-property}
\begin{tabular}{|l|l|c|c|c|c|c|c|}
\hline
\multicolumn{2}{|c|}{Selected AI areas}                                     & Privacy & Stability & Fairness & Security & Composition & Utility              \\ \hline
\multirow{3}{10mm}{Machine learning}   & Private learning             & Yes     &           &          &          & Yes         & Decrease             \\ \cline{2-8}
                                    & Stability in learning        &         & Yes       &          &          &             & Increase             \\ \cline{2-8}
                                    & Fairness in learning         &         &           & Yes      &          &             & Increase             \\ \hline
\multirow{3}{10mm}{Deep learning}      & Deep Learning                & Yes     &           &          &          &             & Decrease             \\ \cline{2-8}
                                    & Distributed deep learning    & Yes     &           &          &          & Yes         & Decrease             \\ \cline{2-8}
                                    & Federated learning           & Yes     &           & Yes      &          & Yes         & Decrease or Increase \\ \hline
\multirow{3}{10mm}{Multi-agent system} & Reinforcement learning       & Yes     &           &          & Yes      & Yes         & Increase             \\ \cline{2-8}
                                    & Auction                      & Yes     &           &          &          & Yes         & Decrease             \\ \cline{2-8}
                                    & Game theory                  &      &           &          &          & Yes         & Decrease             \\  \hline
\end{tabular}
\end{table*}

Table~\ref{table-property} shows the properties that have been explored to date for each of our three disciplines. In machine learning, differential privacy has been applied to private learning, stability and fairness. In deep learning, privacy is the major concern, but distributed deep learning and federated learning have also been investigated. In multi-agent systems, differential privacy has been used to guarantee privacy, provide security, and ensure composition. Utility shows the ultimate performance of the technology after adding differential privacy. Normally, privacy-preserving comes with a utility cost. However, if the differential privacy can contribute to stability, or security, the utility may increase, such as in federated learning or fairness.

Note, however, that blank cells do not mean we can not apply differential privacy mechanisms to those areas. As differential privacy has been proved to work well in many AI areas, in the future, more problems might be solved with the advantages of differential privacy.

The purpose of this paper is to highlight several possible avenues to integrate AI with differential privacy mechanisms, showing that differential privacy mechanisms have several attractive properties that make it quite valuable as a tool to AI beyond merely preserving privacy. The contributions of the paper are listed as follows:
\begin{itemize}
    \item We have summarized several properties of differential privacy mechanisms.
    \item We have shown that these properties can improve diverse aspects of AI areas, including machine learning, deep learning and multi-agent systems.
    \item We explored new possibilities for taking advantage of differential privacy to bring new opportunities.
\end{itemize}

\section{Preliminary}\label{sec:preliminary}
\subsection{Differential privacy}

Consider a finite data universe $\mathcal{X}$.
Let the variable $r$ represent a record
with $d$ attributes sampled from the universe $\mathcal{X}$,
a dataset $D$ is
an unordered set of $n$ records from domain $\mathcal{X}$.
Two datasets $D$ and $D'$ are neighbouring datasets if
they differ in only one record.
A query $f$ is a function that
maps a dataset $D$ to an abstract range $\mathbb{R}$:
$f: D\rightarrow\mathbb{R}$.

The target of differential privacy is to mask the differences between the results to query $f$
between the neighbouring datasets to preserve privacy.
The maximal difference
is defined as the sensitivity $\Delta f$,
which determines how much perturbation is required for a private-preserving answer.
To achieve this goal,
differential privacy provides
a mechanism $\mathcal{M}$,
which is a randomization algorithm that
accesses the database and implements some functionalities.
A formal definition of differential privacy follows.

\begin{definition}[$\epsilon, \delta$-Differential Privacy~\cite{Dwork2006}]\label{Def-DP}
A randomized algorithm $\mathcal{M}$ gives $\epsilon$-differential privacy for any pair of
\emph{neighbouring datasets} $D$ and $D'$,
and, for every set of outcomes $\Omega$,
if $\mathcal{M}$ satisfies:
\begin{equation}
Pr[\mathcal{M}(D) \in \Omega] \leq \exp(\epsilon) \cdot Pr[\mathcal{M}(D') \in \Omega]+\delta
\end{equation}
where $\Omega$ denotes the output range of the algorithm $\mathcal{M}$.
\end{definition}

In Definition~\ref{Def-DP},
the parameter $\epsilon$ is defined as the privacy budget,
which controls the privacy guarantee level of mechanism $\mathcal{M}$.
A smaller $\epsilon$ represents stronger privacy.
If $\delta=0$, the randomized mechanism $\mathcal{M}$
gives $\epsilon$-differential privacy
by its strictest definition.
$(\epsilon,\delta)$-differential privacy provides freedom to
violate strict $\epsilon$-differential privacy
for some low probability events.

Sensitivity is a parameter used in both mechanisms to determine how much randomization is required:
\begin{definition}[Sensitivity]\label{Def-GS}
For a query $f:D\rightarrow\mathbb{R}$,
the sensitivity of $f$ is defined as
\begin{equation}
\Delta f=\max_{D,D'} ||f(D)-f(D')||_{1}
\end{equation}
\end{definition}

Two prevalent randomization mechanisms, Laplace and exponential, are used to satisfy the definition of differential privacy, but there are others, such as Gaussian mechanism. Each is explained next.

\subsection{Randomization: Laplace mechanism}

The Laplace mechanism is applied to numeric outputs~\cite{DworkCalibrated}.
The mechanism adds independent noise to the original answer, as shown in Definition~\ref{Def-LA}.

\begin{definition}[Laplace mechanism]\label{Def-LA}
For a function $f: D \rightarrow \mathcal{R}$ over a dataset $D$,
the mechanism $\mathcal{M}$ in Eq.~\ref{eq-lap} provides $\epsilon$-differential privacy.
\begin{equation}
\mathcal{M}(D)=f(D)+Lap(\frac{\Delta f}{\epsilon})
\end{equation}\label{eq-lap}
\end{definition}

\subsection{Gaussian mechanism}

Compared to a Laplace mechanism,
a Gaussian mechanism adds noise that is sampled from a zero-mean isotropic Gaussian distribution.
The noise $Z$ is sampled $\sim\mathcal{N}(0,\sigma^2)$ to the $L_{2}$ sensitivity
$\Delta f=\max_{D,D'} ||f(D)-f(D')||_{2}$ as follows:

\begin{definition}[Gaussian mechanism]\label{Def-GS}
For a function $f: D \rightarrow \mathcal{R}$ over a dataset $D$,
the mechanism $\mathcal{M}$ in Eq.~\ref{eq-gau} provides $\epsilon, \delta$-differential privacy.
\begin{equation}
\mathcal{M}(D)=f(D)+\sim\mathcal{N}(0,\sigma^2),
\end{equation}\label{eq-gau}
$\sigma=\Delta f \sqrt{2\log(1.25/\delta)/epsilon}$.
\end{definition}

\subsection{Exponential mechanism}

Exponential mechanisms are used to randomize the results for non-numeric queries. They are paired with a score function $q(D,\phi)$
that evaluates the quality of an output $\phi$.
Defining a score function is application-dependent, so
different applications lead to various score functions~\cite{McSherry07STOC}.

\begin{definition}[Exponential mechanism]\label{Def-EX}
Let $q(D,\phi)$ be a score function of dataset $D$
that measures the quality of output $\phi\in\Phi$,
$\Delta q$ represents the sensitivity of $\phi$.
The exponential mechanism $\mathcal{M}$ satisfies $\epsilon$-differential privacy if
\begin{equation}
\mathcal{M}(D) = \left( \text{return $\phi$ }\propto
\exp (\frac{\epsilon q(D,\phi)}{2\Delta q})\right).
\end{equation}
\end{definition}

\subsection{Composition}
Two privacy budget composition theorems are widely used
in the design of differential privacy mechanisms:
sequential composition~\cite{McSherry07STOC} and parallel composition~\cite{McSherry201089}.

\begin{thm}\label{def-comp1}
\emph{Parallel Composition}:
Suppose we have a set of privacy steps
$\mathcal{M}=\{\mathcal{M}_1,...\mathcal{M}_m\}$,
if each $\mathcal{M}_i$ provides an $\epsilon_{i}$ privacy guarantee on
a disjointed subset of the entire dataset,
the parallel of $\mathcal{M}$ will provide
$\max\{\epsilon_{1},...,\epsilon_{m}\}$-\emph{differential privacy}.
\end{thm}
Parallel composition corresponds to cases where
each $\mathcal{M}_i$ is applied to disjointed subsets of the dataset.
The ultimate privacy guarantee only depends on the largest privacy budget.

\begin{thm}\label{def-comp2}
\emph{Sequential Composition}:
Suppose a set of privacy steps
$\mathcal{M}=\{\mathcal{M}_1,...\mathcal{M}_m\}$
are sequentially performed on a dataset,
and each $\mathcal{M}_i$ provides an $\epsilon$ privacy guarantee,
$\mathcal{M}$ will provide $(m\cdot\epsilon)$-\emph{differential privacy}.
\end{thm}
Sequential composition
offers a privacy guarantee for a sequence of differentially private computations.
When a series of randomized mechanisms are performed sequentially on a dataset,
the privacy budgets are added up for each step.

%These two composition theorems bound the degradation of privacy
%when composing several differentially private mechanisms.
%They are the most prevalent and straightforward ways to analyze the privacy budget’s
%consumption of a privacy-preserving algorithm.

\section{Differential Privacy in Machine Learning}\label{sec:DP-ML}

%The aim of machine learning is to develop algorithms
%that can learn how to perform certain tasks
%based on data \cite{Rubaie19}.
%The task could give accurate predictions or finding structures in data.
%However, as we mentioned in Figure~\ref{Fig-examplelearning}, some urgent problems,
%such as privacy violations,
%the stability of learning and fairness in models, need to be tackled.

\subsection{Private machine learning}

Private machine learning aims to protect the individual’s privacy in training data or learning models.
Differential privacy has been considered to be one of the most important tools
in private machine learning and has been heavily investigated in the past decade.

The essential mechanisms in differential privacy all work to extend
current non-private machine learning algorithms into differentially private algorithms.
These extensions can be realized by incorporating Laplace or exponential mechanisms
into non-private learning algorithms directly~\cite{Zhu17survey},
or by adding Laplace noise into the objective functions~\cite{Chaudhuri20111069}.

Starting with Kasiviswanathan et al.'s work~\cite{PrivateLearning08},
the line of research presenting the details of private learning process from privacy based on empirical risk minimization \cite{Chaudhuri11,Wang17}, to prediction \cite{Dwork18,Dagan19},
Bayesian inference \cite{Foulds16,Bernstein18,Bernstein19} and the multi-armed bandit~\cite{Tossou16,Tossou17}.
%Also, private machine learning has been applied to a variety of applications, such as recommendation systems~\cite{Zhu14FGCS,Meng18}, Cyber Physical system~\cite{ZHU2020}, etc.

Private machine learning is one of the most powerful models accepted in this field. To avoid redundancy, this paper will not dive into details of private machine learning.
A number of survey papers have discussed this field~\cite{Zhu17survey,Moha17,Rubaie19,Jay19} thoroughly.

\subsection{Differential privacy in learning stability}

\subsubsection{The overview of stability of learning}

A stable learning algorithm is one in which the prediction does not change much when the training data is modified slightly.
Bousquet et al.~\cite{bousquet2002stability} have proved that stability is linked to the
generalization error bound of the learning algorithm,
indicating that a highly stable algorithm leads to a less over-fit result.
However, increasing the stability of the algorithm is challenging when the size of the testing data is limited.
This is because the validate data sometimes are reused and lead to an incorrect learning model.
%Testing datasets are not reusable.
%If the outcome of the validation is reused
%to select an additional data statistic, this reuse of testing data
%will lead to an incorrect statistical inference.
To preserve statistical learning validity, analysts
should collect new data for a fresh testing set.

Differential privacy can be naturally linked to learning stability.
The concept of differential privacy ensures that the probability
of observing any outcome from an analysis is
essentially unchanged by modifying any single record.
Dwork et al.~\cite{Dwork2015STOC, Dwork14Science} showed that differential
privacy mechanisms can be used to develop adaptive data analysis
algorithms with provable bounds for over-fitting, noting that
certain stability notions
are necessary and sufficient for generalization.
Therefore,
differential privacy is stronger
than previous notions of stability
and, in particular, possesses strong adaptive composition
guarantees~\cite{Hardt2014FOCS}.

\subsubsection{Differential privacy in learning stability}
Dwork et al.~\cite{Dwork14Science} show that, by adding noise to generate fresh testing data, differential privacy mechanisms can achieve highly stable learning.
For a dataset $D$, an analyst learns about the data by running a series of analyses $f_{i}$ on the dataset.
The choice of which analysis to run depends on
the results from the earlier analyses.
Specifically,
the analyst first selects a statistic $f_{0}$ to query on $D$ and observes a query result $y_{1}=f_{0}(D)$.
From the $k^{th}$ analysis, the analyst selects a function $f_{k}$ based on the query result ${y_1,...,...y_{k-1}}$.
To improve the generalization capability of the adaptive scenario,
noise is added in each analysis iteration.
For example, $y_{k}=Lap(\frac{\Delta f}{\epsilon})+f_{k-1}(D)$~\cite{Dwork2015STOC}.

This type of adaptive analysis can be linked to machine learning.
The dataset $D$ can be randomly partitioned into
a training set $D_{t}$ and a testing (holdout) set $D_{h}$.
The analyst can access
the training set $D_{t}$ without restrictions but
may only access $D_{h}$ through a differentially private interface.
The interface takes the testing and training
sets as inputs and, for all functions given by the
analyst, provides statistically valid estimates of
each function’s results.

%Multiplicative weights update mechanism~\cite{NIPS2012} can be applied in this
%interface to save the privacy budget.
%Given a function $f$, the algorithm first checks
%if the difference between the average value of $f$
%on the training set $D_{t}$  and
%the average value of $f$ on the testing set $D_{h}$
%is below a certain threshold
%$T+\eta$. If the difference is below the
%threshold, then the algorithm returns $f(D_{t})$;
%otherwise, the algorithm
%returns $f(D_{h})+Lap(\frac{\Delta f}{\epsilon})$.

For a sufficiently large testing set, the differential privacy interface
guarantees that for function $f:D\rightarrow[0,1]$,
the mechanism will return a randomized value $v_{f}$.
When $v_{f}$ is compared to the query result $y$,
we have $|v_{f}-y|\leq\tau$ with a probability of at least
$1-\beta$, where $\tau$ is the analyst’s choice of error and $\beta$ is the confidence parameter.
The probability space is over the data elements in $D_{h}$ and $D_{t}$ and the randomness
introduced by the interface.
A multiplicative weight updating mechanism~\cite{NIPS2012} could also be included in the
interface to conserve the privacy budget.

\subsubsection{Summary of stability of learning}
The idea of adding randomization during data analysis to increase stability has been widely accepted.
MacCoun et al.~\cite{MacCoun15Nature}
believed: when deciding which results to report, the analyst interacts
with a dataset that has been obfuscated
through adding noise to observations, removing some data points, or switching data labels.
The raw, uncorrupted, dataset is only used in computing the final reported values.
Differential privacy mechanisms can follow the above rules to significantly improve learning stability.

\subsection{Differential privacy in fairness}

\subsubsection{An overview of the fairness in learning}

Fairness issues are prevalent in every facet of our lives
including education, job application, the parole of prisoners and so on \cite{Binns18,Holstein19,Zhang20}.
Instead of resolving fairness issues, modern AI techniques, however, can amplify social inequities and unfairness.
For example, an automated hiring system may be more likely to recommend candidates from specific racial, gender or age groups \cite{Boettcher17,Giang18}.
A search engine may amplify negative stereotypes by showing arrest-record ads
in response to queries for names predominantly given to African-American babies but not for other names \cite{BBC13,Nobel18}.
Moreover, some software systems that are used to measure the risk of a person recommitting crime demonstrate a bias against African-Americans over Caucasians with the same profile~\cite{Angwin16,Mehrabi19}.
To address these fairness issues in machine learning, great effort has been placed on developing definitions of fairness \cite{Zemel13,Hardt16,Dwork19}
and algorithmic methods for assessing and mitigating undesirable bias in relation to these definitions~\cite{Kusner17,Agarwal18b}.
A typical idea is to make algorithms insensitive to one or multiple attributes of datasets, such as gender and race.

\subsubsection{Applying differential privacy to improve fairness}

Dwork et al.~\cite{Dwork12} classified individuals with the goal of preventing discrimination against a certain group while maintaining utility for the classifier. The key idea is to treat similar individuals similarly. To implement this idea, these researchers adopted the Lipschitz property, which requires that any two individuals,
$x$ and $y$, with a distance of $d(x,y)\in[0,1]$
must map to the distributions $M(x)$ and $M(y)$, respectively, such that the statistical distance
between $M(x)$ and $M(y)$ is at most $d(x,y)$~\cite{dixit2013testing}.
In other words, if the difference between $x$ and $y$ is $d(x,y)$,
the difference of the classification outcomes of $x$ and $y$ is at most $d(x,y)$.
A connection between differential privacy and the Lipschitz property has been theoretically established, in that a mapping satisfies differential privacy if, and only if, this mapping satisfies the Lipschitz property~\cite{Dwork12}.

Zemel et al.~\cite{zemel13pmrl} extended Dwork et al.’s~\cite{Dwork12} preliminary work by defining the metrics between individuals. They learned a restricted form of a distance function and formulated fairness as an optimization problem of finding the intermediate representation that best encodes the data. During the process, they preserved as much information about the individual’s attributes as possible, while removing any information about membership with other protected subgroup. The goal was two-fold: first, the intermediate representation should preserve the data’s original features as much as possible. Second, the encoded representation is randomized to hide whether or not the individual is from the protected group.

Both ideas take advantage of randomization in differential privacy. Considering an exponential mechanism with a carefully designed score function, the framework can sample fresh data from the universe to represent original data with the same statistical properties. However, the most challenging part of this framework is designing the score function. This is because differential privacy in fairness assumes that the similarity between individuals is given; however, estimating similarity between individuals in an entire feature universe is a tough problem. In other words, the evaluation of similarity between individuals is the key obstacle of model fairness, making score function design an obstinate problem. Therefore, differential privacy in model fairness needs further exploration.

Recently, researchers have attempted to adopt differential privacy to simultaneously achieve both fairness and privacy preservation~\cite{Xu19WWW,Ding20}. This research is motivated by settings where models are required to be non-discriminatory in terms of certain attributes, but these attributes may be sensitive and so must be protected while training the model~\cite{Jagielski19}. Addressing fairness and privacy preservation simultaneously is challenging because they have different aims~\cite{Cum19,Ding20}. Fairness focuses on the group level and seeks to guarantee that the model’s predictions for a protected group (such as female) are the same as the predictions made for an unprotected group. In comparison, privacy preservation focuses on the individual level. Privacy preservation guarantees that the output of a classification model is independent of whether any individual record is in or out of the training dataset. A typical solution to achieve fairness and privacy preservation simultaneously was proposed by Ding et al.~\cite{Ding20}.
Their solution is to add a different amount of differentially private noise based on different polynomial coefficients of the constrained objective function, where the coefficients relate to attributes in the training dataset. Therefore, privacy is preserved by adding noise to the objective function, and fairness is achieved by adjusting the amount of noise added to the different coefficients.

\subsubsection{Summary of differential privacy in fairness}

The best methods of similarity measurement and composition are open problems in fairness models. Further, differential privacy in fairness models has been directed toward classification problems. There are also some works on fairness in online settings such as online learning, bandit learning and reinforcement learning. However, how to use differential privacy mechanisms to benefit those online settings of fairness needs further investigation.

Composition fairness is also a big challenge. Here, fairness means that if each component in the algorithm satisfies the notion of fairness, the entire algorithm will satisfy the same~\cite{Alexandra18}. This composition property is essential for machine learning, especially for online learning. Dwork et al.~\cite{Dwork18Composition} explored this direction, finding that current methods seldom achieve this goal because classification decisions cannot be made independently, even by a fair classifier. Also, classifiers that satisfy group fairness properties may not compose well with other fair classifiers. Their results show that the signal provided by group fairness definitions under composition is not always reliable. Hence, further study is needed to figure out how to take advantage of differential privacy to ensure composition.

\subsection{Summary of differential privacy in machine learning}

\begin{table*}[htp]\scriptsize
\newcommand{\tabincell}[2]{\begin{tabular}{@{}#1@{}}#2\end{tabular}}
	\centering
	\caption{Summary of differential privacy in machine learning}
	\scalebox{0.9}{
\begin{tabular}{|c|c|c|c|c|c|} \hline
\textbf{Papers}&\textbf{Research areas}&\textbf{Techniques used}&\textbf{Research aims}&\textbf{Advantages}&\textbf{Disadvantages}\\ \hline
Dwork et al.~\cite{Dwork14Science,Dwork2015STOC} & \tabincell{c}{Stable learning} & \tabincell{c}{Laplace \\mechanism} & Improve stability & \tabincell{c}{Improve stability \\with little overhead} & \tabincell{c}{Limited access \\to testing dataset}\\ \hline
Dwork et al. \cite{Dwork12} & \tabincell{c}{Fairness in learning} & \tabincell{c}{Concept of \\differential privacy} & Improve fairness & \tabincell{c}{Not only enforce\\ fairness but also\\ detect unfairness} & \tabincell{c}{An available similarity \\metric is assumed \\a prerequisite}\\ \hline
Zemel et al. \cite{zemel13pmrl} & \tabincell{c}{Fairness in learning} & \tabincell{c}{Concept of \\differential privacy} & Improve fairness & \tabincell{c}{Simultaneously encode \\and obfuscate data} & \tabincell{c}{Representation dependent} \\ \hline
Xu et al. \cite{Xu19WWW} & \tabincell{c}{Fairness in learning} & \tabincell{c}{Laplace \\mechanism} & \tabincell{c}{Improve fairness and \\preserve privacy} & \tabincell{c}{Achieve both fairness \\and privacy} & \tabincell{c}{For logistic \\regression only}\\ \hline
Jagielski et al. \cite{Jagielski19} & \tabincell{c}{Fairness in learning} & \tabincell{c}{Laplace \\mechanism} & \tabincell{c}{Improve fairness and \\preserve privacy} & \tabincell{c}{Achieve both \\fairness and privacy} & \tabincell{c}{Need large dataset}\\ \hline
Ding et al. \cite{Ding20} & \tabincell{c}{Fairness in learning} & \tabincell{c}{Functional \\mechanism} & \tabincell{c}{Improve fairness and \\preserve privacy} & \tabincell{c}{Achieve both fairness \\and privacy} & \tabincell{c}{For logistic \\regression only}\\ \hline
\end{tabular}}
	\label{tab:summaryML}
\end{table*}

Table \ref{tab:summaryML} summarizes the papers that apply differential privacy to learning stability and fairness. From this summary, we can see that differential privacy can not only preserve privacy but also improve the stability and fairness in machine learning.
The key idea of achieving stability is derived from allowing an analyst to access the testing set only in a differentially private manner.
Likewise, the main idea of achieving fairness is also derived from randomly re-sampling fresh data from the data universe in a differentially private manner.
The two examples show that the sampling from the data universe can improve the machine learning performance to some extent.
% Add the key idea of stability is derived from the randomization, and fairness is derived from re-sampling.

Even though differential privacy has been proven to
guarantee privacy, stability and fairness in machine learning,
there are still some open research issues.
First, to preserve privacy, the utility of learning models is sacrificed to some extent.
Thus, how to obtain an optimal trade-off between the privacy and the utility still needs further exploration.
Second, current differentially private stable learning is suitable only for the learning models
where loss functions do not have regularization.
Differential privacy can provide additional generalization capability to the learning models who has limited regularization.
Hence, improving generalization capability for regularized loss functions will be helpful.
Third, the re-sampling in current fair learning is typically based on the exponential mechanism.
Exponential mechanism requires the knowledge of the utility of each sample.
This knowledge, however, may not be available or hardly be defined in some situations.
Thus, new mechanisms are needed for today's fair learning.

Research on differential privacy in machine learning can be broadened to address other non-privacy issues. For example, differential privacy mechanisms may be able to generate new data samples based on existing ones by properly adding noise to the values of attributes in existing samples. These newly-generated samples may not be suitable for training data, but they can be used as testing data. Another example is that differential privacy mechanisms can be used for sampling. Sampling is an important step in deep reinforcement learning and batch learning. The small database mechanism may be a good tool for sampling in machine learning, as it can guarantee the desired accuracy while sampling only a small set of samples.

% add a conclusion and future development of dp in machine learning.
% Currently, even dp has been proved to gurantee stability, fairness, privacy in machine learning, there are still open problems. For privacy, the utility is still an issue. In stability, actually, it suitable for the learning that the loss function without any regularization. In another word, the generalization capability is not quite strong. For the fairness, re-sampling is based on the exponential mechanism, but can be extended to more mechanisms.

\section{Differential privacy in deep learning}

Deep learning originated from regular neural networks
but thanks to the availability of large volumes of data and advancements in computer hardware,
implementing many-layered neural network models has become feasible,
and these models significantly outperforms their predecessors.
%\cite{Pouy18}.
%In 2016/2017, Google's AlphaGo, trained with deep learning models, defeated the top human Go players \cite{AlphaGo,Mnih15}.
%This great success skyrocketed deep learning into becoming one of the most important research topics in machine learning,
%and subsequently, a great many of deep learning algorithms have been proposed \cite{Alom19,Jiao19}.
The latest deep learning algorithms have been successfully applied to many applications
such as natural language processing, image processing, and speech and audio processing \cite{Dargan19}.
%Training a deep learning model usually requires a large amount of data.
%However, these data may contain private information.
Differential privacy has been broadly used in deep learning to preserve data and model privacy.
Thus, in this section, we mainly focus on analyzing the differential privacy in general deep neural networks, distributed deep learning \cite{Pouy18} and federated learning \cite{Yang19}.
%Distributed deep learning and federated learning allow a set of independent data owners
%to collaboratively learn a model based on their data sets without disclosing their private data \cite{Jay18}.
%In the following sub-sections, we first provide an overview of differential privacy in general deep neural networks
%and then thoroughly review differential privacy in distributed deep learning and federated learning.

\subsection{Deep neural networks: attacks and defences}

\subsubsection{Privacy attacks in the deep neural networks}
One of the most common privacy attacks is an inference attack
where the adversary maliciously infers sensitive features and background information
about a target individual from a trained model \cite{Gong20}.
%The deep neural network has achieved great success in modern artificial intelligence applications.
%The training process of deep neural networks, however, may reveal individual privacy.
%Based on the trained model and background information of a target individual,
%the adversary can maliciously infer the sensitive feature of the target individual \cite{Gong20}.
Typically, there are two types of inference attacks in deep learning.

The first type is a membership inference attack.
The aim here is to infer whether or not a given record is in the present training dataset \cite{Nasr19}.
Membership inference attacks can be either black-box or white-box.
Black-box means that an attacker can query a target model
but does not know any other details about that model, such as its structure \cite{Shokri17}.
In comparison, white-box means that an attacker has full access to a target model along with some auxiliary information \cite{Yeom18}.
%Formally, in a membership inference attack, the attacker is given a data sample $z=(x,y)$
%and has access to a target model $T$, which is trained on dataset $D$.
%The attacker also knows the distribution from which the training set $D$ has been drawn.
%With this information, the aim of the attacker is to decide whether $z\in D$.
%A typical membership inference attack model was proposed by Shokri et al. \cite{Shokri17}.
The attack model is based on the observation that
machine learning models often behave differently on training data versus
the data they ``see'' for the first time.
%This attack model $A$ is implemented by machine learning algorithms
%whose input is a data sample $z$ and the output is binary: $in$ and $out$
%which indicates $z\in D$ and $z\notin D$, respectively.
%To train this attack model, the attacker creates a collection of $k$ shadow models $\{Sh_1,...,Sh_k\}$.
%Each shadow model $Sh_i$ is trained on a synthetic dataset $D^{train}_{Shadow^i}$
%which has the same format as and distributed similarly to the target model's training dataset $D$.
%During the training, for each shadow model $Sh_i$, for all the samples in the training set of shadow model $Sh_i$,
%$\forall (x,y)\in D^{train}_{Shadow^i}$, compute the prediction vector $\mathbf{y}=Sh_i(x)$
%where $\mathbf{y}$ is a probability distribution over the possible classes of $x$.
%Then, the record $(y,\mathbf{y},in)$ is added into the training set $D^{train}_{Attack}$ of attack model $A$.
%Moreover, for all the samples in the test set of shadow model $Sh_i$, $\forall (x,y)\in D^{test}_{Shadow^i}$,
%compute the prediction vector $\mathbf{y}=Sh_i(x)$
%and add the record $(y,\mathbf{y},out)$ into the training set $D^{train}_{Attack}$.
%Finally, the attacker splits $D^{train}_{Attack}$ into $|\mathbf{y}|$ partitions,
%each associated with a different class label.
%Then, for each label $y$, the attacker trains a separate model
%which predicts the $in$ or $out$ status of $x$ given the vector $\mathbf{y}$.}

The second type of attack is an attribute inference attack.
The aim of an attribute inference attack is to learn hidden sensitive attributes of a test input
given access to the model and information about the non-sensitive attributes \cite{Fredrikson14}.
%Formally, in an attribute inference attack, the attacker is given a data sample $z=(\mathbf{x},y)$, where %%$\mathbf{x}=(\mathbf{x}_v,\mathbf{x}_t)$, and
%$\mathbf{x}_v$ are the values of non-sensitive attributes known to the attacker and
%$\mathbf{x}_t$ are the sensitive attributes targeted in the attack.
%The attacker also has access to a target model $T$ trained on dataset $D$,
%and knows the distribution from which the training set $D$ is drawn.
%With this information, the aim of the attacker is to infer the probability that
%$\mathbf{x}_t$ is of a certain value $t$, i.e., $Pr[\mathbf{x}_t=t|\mathbf{x}_v,y]$.
Fredrikson \cite{Fredrikson14} describes a typical attribute inference attack method,
which attempts to maximize the posterior probability estimate of the sensitive attributes
based on the values of non-sensitive attributes.
%The attacker first generates a dataset $D'=\{(\mathbf{x}',y)|\mathbf{x}'_v=\mathbf{x}_v\wedge T(\mathbf{x}')=y\}$
%which is drawn from the same distribution as the training dataset $D$.
%The attacker then can compute $Pr[\mathbf{x}_t=t|\mathbf{x}_v,y]$ using Equation \ref{eq:attrAttack}:
%\begin{equation}\label{eq:attrAttack}
%    Pr[t|\mathbf{x}_v,y]=\frac{Pr[t,\mathbf{x}_v,y]}{Pr[\mathbf{x}_v,y]}=\frac{\sum_{(\mathbf{x}',y)\in D':\mathbf{x}'_t=t}p(\mathbf{x}',y)}{\sum_{(\mathbf{x}',y)\in D'}p(\mathbf{x}',y)}.
%\end{equation}
%In the case that the attacker is unaware of the underlying joint prior $p$,
%he can apply the principal of maximum entropy \cite{Homer08} to estimate $p$.

\subsubsection{Differential privacy in deep neural networks}

\begin{figure}[ht]
\centering
	\includegraphics[scale=0.65]{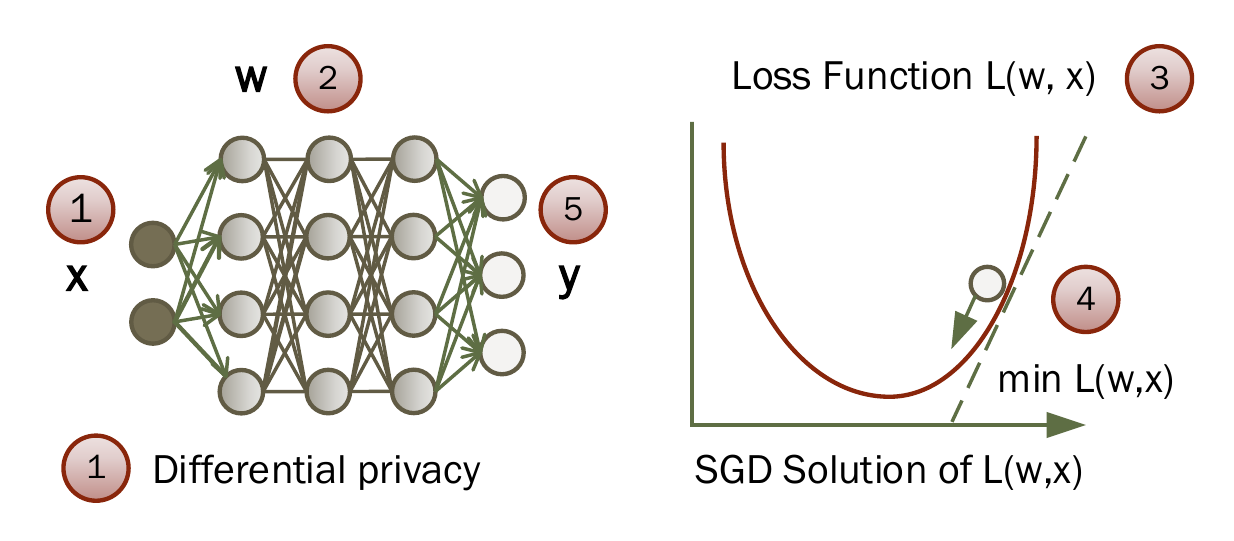}
	\caption{Differential privacy in deep learning}
	\label{fig-dpattackdefend}
\end{figure}

%Although differential privacy was originally introduced into deep learning
%to offer protection against private and sensitive information leaks from deep neural networks \cite{Jay19},
Some of the properties of differential privacy are naturally resistant to membership and attribute inference attacks.
%For membership inference attacks, the aim of an attacker is to determine whether a data sample $z$ is in the training set $D$.
%Thus, to resist membership inference attacks, given a learned model $T$, we expect that $T(D)\approx T(D')$,
%where $D$ and $D'$ differ in a data sample $z$.
%One of the original goals of differential privacy is to mask the difference between the query outputs from two neighbouring datasets.
%For attribute inference attacks, the aim of an attacker is to infer the values of sensitive attributes in the training dataset.
An intuitive way to resist inference attacks is to properly add differentially private noise to the values of the sensitive attributes before using the dataset to train a model.
%Similarly, adding differential private noise to deep learning algorithms
%can create algorithms that satisfy differential privacy.
In a typical deep learning algorithm, there are four places to add noise,
as shown in Figure~\ref{fig-dpattackdefend}.
The first place is the training dataset, where the noise is derived from an input perturbation mechanism.
%As mentioned above, the noise can be added to the values of sensitive attributes to protect their privacy.
This operation occurs before the training starts
and is usually done to resist attribute inference attacks.
%The operation is easy to implement
%but the values of the parameters in differential privacy must be carefully adjusted
%to achieve a tradeoff between the utility of the learned model and the privacy of the training dataset.

The second place is the loss function, which yields an objective perturbation mechanism.
%A loss function is defined as the difference between the ground truth and the learned result.
%The training goal is to ensure this difference as small as possible without overfitting.
%Yet overfitting is one of the theoretical foundations of membership inference attacks.
%Adding noise to the loss function can avoid overfitting to some extent by randomly obfuscating the difference between the ground %truth and the learned result.
This operation occurs during training
and is usually done to resist membership inference attacks.

The third place is the gradients at each iteration, i.e., a gradient perturbation mechanism.
Gradients are computed using the loss function to do partial-derivative against the weights of the deep neural network.
%Since noise is typically independent of the weights,
%the effect of adding noise to the gradients is similar to the effect of adding noise to the loss function.
Likewise, this operation occurs during training
and is usually done to resist membership inference attacks.

The fourth place is the weights of the deep neural network, constituting the learned model, called an output perturbation mechanism.
%The weights of the deep neural network constitute the learned model.
This operation happens once training is complete.
The operation is easy to implement
and can resist both membership and attribute inference attacks.
However, directly adding noise to the model may significantly harm its utility,
even if the parameter values in differential privacy have been carefully adjusted.
%A potential way to conserve the utility of the model when using differential privacy is to adopt matrix-based differential privacy \cite{Chan18}
%that can conserve the characteristics of a matrix when modifying the values in the matrix.

Of these four places, adding noise to the gradients is the most common method.
However, because the gradient norm may be unbounded in deep learning,
a bound must be imposed before applying gradient perturbation.
%Application of gradient perturbation needs a bound on the gradient norm,
%as the gradient norm may be unbounded in deep learning.
A typical way is to manually clip the gradient at each iteration \cite{Abadi16}.
Such a clipping can also provide a sensitivity bound with differential privacy.
Table~\ref{tab-dpnoise} summarizes the properties of the mentioned attacks and their defence strategies.

\begin{table*}  \centering
  \caption{Attacks and defense in deep learning}\label{tab-dpnoise}
  \begin{tabular}{c|c|c|c|c}
\toprule
   Noise & Membership inference attack & Attribute inference attack & Privacy guarantee & Performance impact \\
\midrule
    Dataset \cite{Heikkila17} &  & $\checkmark$ & very strong & high \\
    Loss function \cite{Zhao19} & $\checkmark$ & & strong & low \\
    Gradient \cite{Shokri15,Abadi16,Cheng18} & $\checkmark$ & & strong & low \\
    Weights \cite{Jay18,Phan19} & $\checkmark$ & $\checkmark$ & very strong & very high \\
    Classes~\cite{Papernot17, Papernot18,Zhao18} & $\checkmark$ & $\checkmark$ & very strong & low \\
\bottomrule
\end{tabular}
\end{table*}

From Table \ref{tab-dpnoise}, we can see that adding noise to a dataset can defend against attribute inference attacks.
Since the aim of attribute inference attacks is to infer the values of sensitive attributes,
directly adding noise to these values is the most straightforward and efficient method of protecting them.
However, this method may significantly affect the utility of the learned model,
because it is heavily dependent on the values of the attributes in the training dataset,
and using a dataset with modified attribute values to learn a model is similar to using a ``wrong'' dataset.
%The learning algorithm can still learn a model based on the ``wrong'' dataset,
%but the learned model may not be the expected one and have low utility.

By comparison, adding noise to the loss function or gradient only slightly affects the utility of the learned model.
This is because the noise is added during the training process
and the model can be corrected by taking the noise into account.
Adding noise to the loss function or gradient can resist membership inference attacks,
which can be guaranteed by the properties of differential privacy.
However, adding noise to the loss function or gradient does not offer much resistance to attribute inference attacks.
As mentioned before, a typical attribute inference attack needs two pieces of information:
1) the underlying distribution of the training dataset; and 2) the values of non-sensitive attributes.
These two pieces of information are not modified when adding noise to the loss function or gradient.

Finally, adding noise to the weights or classes of a neural network can resist both membership and attribute inference attacks.
This is because adding noise to the weights will modify the learned model and both of these types of attacks need to access the learned model to launch an attack.
The downside is that adding noise to a learned model after training may drastically affect its utility,
and retraining the model will not correct the problem as noise simply needs to be added again.
Noise could be added to the weights in each iteration of training.
However, this method might affect convergence,
since the output of the algorithm is computed based on the weights.
Hence, if noise is added to each weight,
the total amount of noise might become large enough to make the loss never convergent.
Adding noise to the classes has the similar disadvantage for the same reason.

%In the following two sub-sections, we discuss distributed deep learning and federated learning,
%which are both applications of deep neural networks.

\subsection{Differential privacy in distributed deep learning}

%\begin{figure}[ht]
%\centering
%    \includegraphics[scale=0.4]{regress2.png}
%    \caption{Linear regression with [-1,+1] labels}
%\end{figure}

\subsubsection{Overview of distributed deep learning}

Conventional deep learning is limited to a single-machine system,
where the system has all the data and carries out the learning independently.
Distributed deep learning techniques, however, accelerate the learning process.
%To accelerate the learning process, distributed deep learning techniques are developed.
Two main approaches are applied in distributed deep learning:
data parallelism and model parallelism \cite{Pouy18}.
In data parallelism, a central model is replicated by a server and distributed to all the clients.
Each client then trains the model based on her own data.
After a certain period of time, each client summarizes an update on top of the model
and shares the update to the server.
The server then synchronizes the updates from all the clients and improves the central model.
In model parallelism, all the data are processed with one model.
The training of the model is split between multiple computing nodes,
with each computes only a subset of the model.
As data parallelism can intrinsically protect the data privacy of clients,
most research on distributed deep learning focuses on data parallelism.

\subsubsection{Differential privacy in distributed deep learning}
%When using differential privacy in distributed deep learning,
%an intuitive way is to add differential privacy noise to the final model parameters.
%This way, however, can destroy the utility of the learned model
%because the amount of noise may be large enough to break the dependence of these parameters on the training data.
%Instead of adding noise to the parameters, typically,
%differential privacy noise is added to the gradient of a learning model.
As mentioned in the previous subsection, differentially private noise can be added to five places in a deep neural network.
%the input datasets, the loss functions, the gradients or the weights, and the output classes.
The following review is divided into methods based on adding noise.

\emph{Adding noise to input datasets.}
Heikkila et al. \cite{Heikkila17} proposed a general approach for privacy-preserving learning in distributed settings.
Their approach combines secure multi-party communication with differentially private Bayesian learning methods
so as to achieve distributed differentially private Bayesian learning.
In their approach, each client $i$ adds a Gaussian noise to her data
and divides them and the noise into shares.
Each share is then sent to a server.
In this way, the sum of the shares discloses the real value,
but separately they are just random noise.

\emph{Adding noise to loss functions.}
Zhao et al. \cite{Zhao19} proposed a privacy-preserving collaborative deep learning system
that allows users to collaboratively build a collective learning model
while only sharing the model parameters, not the data.
To preserve the private information embodied in the parameters,
a functional mechanism, which is an extended version of the Laplace mechanism,
was developed to perturb the objective function of the neural network.

\emph{Adding noise to gradients.}
Shokri and Shmatikov \cite{Shokri15} designed a system
that allows participants to independently train on their own datasets
and share small subsets of their models' key parameters during training.
Thus, participants can jointly learn an accurate neural network model
without sharing their datasets,
and can also benefit from the models of others to improve their learning accuracy
while still maintaining privacy.

Abadi et al. \cite{Abadi16} developed a differentially private stochastic gradient descent algorithm for distributed deep learning.
At each iteration during the learning, Gaussian noise is added to the clipped gradient to preserve privacy in the model.
In addition, their algorithm also involves a privacy accountant and a moment accountant.
The privacy accountant computes the overall privacy cost during the training,
while the moment accountant keeps track of a bound on the moments of the privacy loss random variable.

Cheng et al. \cite{Cheng18} developed a privacy-preserving algorithm for distributed learning
based on a leader-follower framework,
where the leaders guide the followers in the right direction to improve their learning speed.
For efficiency, communication is limited to leader-follower pairs.
To preserve the privacy of leaders, Gaussian noise is added to the gradients of the leaders' learning models.

\emph{Adding noise to weights.}
Jayaraman et al. \cite{Jay18} applied differential privacy with both output perturbation and gradient perturbation in a distributed learning setting.
With the output perturbation, each data owner combines their local model with a secure computation
and adds Laplace noise to the aggregated model estimator before revealing the model.
With the gradient perturbation, the data owners collaboratively train a global model using an iterative learning algorithm,
where, at each iteration, each data owner aggregates their local gradient within a secure computation
and adds Gaussian noise to the aggregated gradient before revealing the gradient update.

Phan et al. \cite{Phan19} proposed a heterogeneous Gaussian mechanism to preserve privacy in deep neural networks.
Unlike a regular Gaussian mechanism, this heterogeneous Gaussian mechanism can arbitrarily redistribute noise
from the first hidden layer and the gradient of the model to achieve an ideal trade-off between model utility and privacy loss.
To obtain the property of arbitrary redistribution, a noise redistribution vector is introduced
to change the variance of the Gaussian distribution.
Further, it can be guaranteed that, by adapting the values of the elements in the noise redistribution vector,
more noise can be added to the more vulnerable components of the model to improve robustness and flexibility.

\emph{Adding noise to output classes.}
Papernot et al. \cite{Papernot17} developed a model called Private Aggregation of Teacher Ensembles (PATE)
which has been successfully applied to generative adversarial nets (GAN) for privacy guarantees \cite{Jordon19}.
PATE consists 1) an ensemble of $n$ teacher models;
2) an aggregation mechanism; and 3) a student model.
Each teacher model is trained independently on a subset of private data.
%The aggregation mechanism is used by teachers to collectively make a decision, by vote regarding the output class of a given sample.
To protect the privacy of data labels, Laplace noise is added to the output classes, i.e., the teacher votes.
Last, the student model is trained through knowledge transfer from the teacher ensemble with the public data and privacy-preserving labels.
Later, Papernot et al. \cite{Papernot18} improved the PATE model to make it applicable to large-scale tasks and real-world datasets.
%They first replaced the Laplace noise added to the teacher votes with Gaussian noise,
%which is better suited to data-dependent privacy analysis.
%They next modified the aggregation mechanism by introducing confident and interactive aggregators
%that select queries worth answering in a privacy-preserving way.
Zhao \cite{Zhao18} also improved the PATE model by extending it to the distributed deep learning paradigm.
Each distributed entity uses deep learning to train a teacher network on private and labeled data.
The teachers then transfer the knowledge to the student network at the aggregator level in a differentially-private manner
by adding Gaussian noise to the predicted output classes of the teacher networks.
This transfer uses non-sensitive and unlabeled data for training.

\subsubsection{Summary of differential privacy in distributed deep learning}
% to do
Although a number of privacy-preserving methods have been proposed for distributed deep learning,
there are still some challenging issues that have not yet been properly addressed.
The first issue is synchronization.
If data parallelism has too many training modules,
it has to decrease the learning rate to ensure a smooth training procedure.
Similarly, if model parallelism has too many segmentations,
the output from the nodes will reduce training efficiency \cite{Pouy18}.
Differential privacy offers potential for solving this issue.
Technically, the challenge is a coordination problem,
where modules or nodes collaboratively perform a task,
but each has a privacy constraint.
This coordination problem can be modeled as a multi-player cooperative game,
and differential privacy has been proven as effective for achieving equilibria in this game \cite{Kearns14}.

The second issue is collusion.
Most of the existing methods assume non-collusion between multiple computation parties.
This assumption, however, may fail in some situations.
For example, multiple service providers may collude to obtain a user's private information.
Joint differential privacy may be able to address this issue,
as it has been proven to successfully protect any individual user's privacy
even if all the other users collude against that user \cite{Zhang16}.

The third issue is privacy policies.
Most existing methods rely on privacy policies
that specify which data can be used by which users according to what rules.
However, there is no guarantee that all the users will strictly follow the privacy policies.
Differential privacy may be available to deal with this issue,
as differential privacy can guarantee users will truthfully report their types
and faithfully follow the recommendations given by the privacy policies.

\subsection{Differential privacy in federated learning}

\subsubsection{Overview of federated learning}

Federated learning enables individual users to collaboratively learn a shared prediction model
while keeping all the training data on the users' local devices.
Federated learning was first proposed by Google in 2017 as an additional approach to the standard centralised machine learning approaches~\cite{Google17},
and has been applied to several real-world applications~\cite{FLsurvey2019}.

\begin{figure}
\centering
	\includegraphics[scale=0.7]{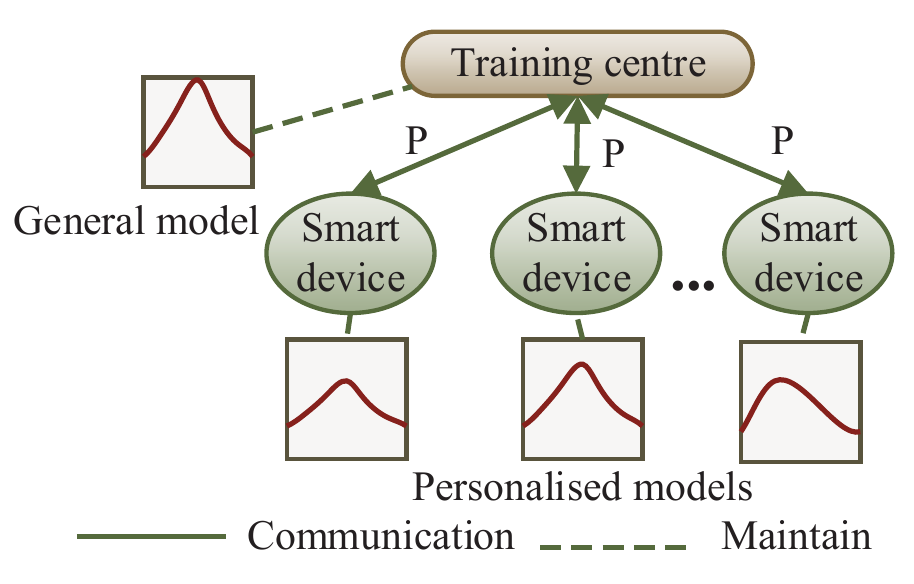}
	\caption{Federated Learning Framework}
	\label{fig-federated}
\end{figure}

Fig.~\ref{fig-federated} shows the structure of a simple federated learning framework.
First, the training centre distributes a general learning model,
trained on general data, to all the smart devices in the network. This model is used
for general purposes, such as image processing.
In a learning iteration, each user downloads a shared model from the training centre to their local device.
Then, they improve the downloaded model by learning from their own local data.
Once complete, the changes to each user's local model are summarized as a small update,
which is sent to the training centre through a secure communication channel.
Once the cloud server receives all the updates,
the shared model is improved wholesale.
During the above learning process, each user's data remain on their own device
and is never shared with either the training centre or another user.

However, although private user data cannot be directly obtained by others,
it may be indirectly leaked through the updates.
For example, the updates of the parameters in an optimization algorithm, such as stochastic gradient descent \cite{McMahan17},
may leak important data information when exposed together with data structures \cite{Phong18}.
Differential privacy can, however, resolve this problem, as explained next.

\subsubsection{Applying differential privacy in federated learning}

Although no training data is transferred from mobile devices to the cloud centre in federated learning,
simply keeping data locally does not provide a strong enough privacy guarantee
when conventional learning methods are used. For example, adversaries can use differential attacks to discover what data was used during training through the parameters of the learning model~\cite{Geyer17}.
To protect against these types of attacks, several algorithms that incorporate differential privacy have been proposed
that ensure a learned model does not reveal whether the data from a mobile device was used during training~\cite{Federated17}.

Adversaries can also interfere with the messages exchanged between communicating parties,
or they can collude among communicating parties during training to attack the accuracy of the learning outcomes.
To ensure the resulting federated learning model maintains acceptable prediction accuracy, approaches using both differential privacy mechanisms and secure multiparty computation frameworks have been created, providing formal data privacy guarantees~\cite{Fedarated18}.

Geyer et al. \cite{Geyer17} incorporated differential privacy mechanisms into federated learning to ensure that
whether an individual client participates in the training cannot be identified.
This approach protects the entire data of an individual client.
To achieve this aim, in each communication round, a subset of the total clients is randomly selected.
Then, the difference between the central model and each of the selected client's local model is calculated,
and Gaussian noise is added to the difference.

Shi et al. \cite{Shi17} investigated a distributed private data analysis setting,
where multiple mutually distrustful users co-exist and each of them has private data.
There is also an untrusted data aggregator in the setting
who wishes to compute aggregate statistics over these users.
The authors adopted computational differential privacy to develop a protocol,
which can output meaningful statistics with a small total error
even when some of the users fail to respond.
Also, the protocol can guarantee the privacy of honest users,
even when a portion of the users are compromised and colluding.

Agarwal et al. \cite{Agarwal18} combined two aspects of distributed optimization in federated learning:
1) quantization to reduce the total communication cost; and
2) adding Gaussian noise to the gradient
before sending the result to the central aggregator to preserve the privacy of each client.

\subsubsection{Summary of differential privacy in federated learning}

The main advantage of the federated leaning model is that
none of the training data needs be transferred to the cloud centre
which satisfies the basic privacy concerns of mobile device users can be satisfied. However,
federated learning has some unique challenges, mainly in the following three respects:
\begin{itemize}
  \item Issues related to attacks on various vulnerabilities of the federated learning model and the countermeasures to defend against these attacks. For example, adversaries can use differential attacks to determine which mobile users have been included in the learning process~\cite{smith2017federated}; messages can be tampered with; and adversaries can use model poisoning attacks to cause the model to misclassify a set of chosen inputs with high confidence~\cite{bhagoji2018analyzing, pmlrv97}.

  \item Issues related to the learning algorithms, such as the requirements of accuracy, scalability, efficiency, fault-tolerance, etc.~\cite{Google17, KonecnyMYRSB16}.

    \item Issues related to the structure of the federated learning system, including its communication efficiency, the computational and power limitation of the mobile devices, the reliability of the mobile devices and communication system, etc.~\cite{smith2017federated}. This issue can potentially be tackled through the composition property of differential privacy by fixing the privacy budget and forcing all communications to consume that budget.
%the communication overhead can be overcome.
\end{itemize}

To effectively use federated learning in various applications,
we first need to overcome the challenges related to attacks, the system structures, and the learning algorithms.
Therefore, intensive research addressing these challenges will be required in the near future.
The second future development will be to explore the power and benefits of federated learning for both new and existing applications,
especially now that mobile devices are ubiquitous.
The third future development will be the automation of tools
that use federated learning and the emergence of companies providing such services to meet the various needs of business and individual customers.

\subsection{Summary of differential privacy in deep learning}
\begin{table*}[!ht]\scriptsize
\newcommand{\tabincell}[2]{\begin{tabular}{@{}#1@{}}#2\end{tabular}}
	\centering
	\caption{Summary of differential privacy in deep learning}
	\scalebox{0.9}{
\begin{tabular}{|c|c|c|c|c|c|} \hline
\textbf{Papers}&\textbf{Research areas}&\textbf{Techniques used}&\textbf{Research aims}&\textbf{Advantages}&\textbf{Disadvantages}\\ \hline
Papernot et al. \cite{Papernot17} & \tabincell{c}{Deep learning} & \tabincell{c}{Laplace\\ mechanism} & Preserve privacy & \tabincell{c}{Independent of \\learning algorithms} & \tabincell{c}{Suitable only for\\ small-scale tasks}\\ \hline
Papernot et al. \cite{Papernot18} & \tabincell{c}{Deep learning} & \tabincell{c}{Gaussian\\ mechanism} & Preserve privacy & \tabincell{c}{Suitable for \\large-scale tasks} & \tabincell{c}{Need two \\aggregators}\\ \hline
Shokri et al. \cite{Shokri15} & \tabincell{c}{Distributed \\deep learning} & \tabincell{c}{Sparse \\vector technique} & Preserve privacy & \tabincell{c}{Preserve the privacy of \\participants without \\sacrificing the accuracy of \\resulting models} & \tabincell{c}{Vulnerable to Generative \\Adversarial Network\\-based attacks \cite{Hitaj17}}\\ \hline
Abadi et al. \cite{Abadi16} & \tabincell{c}{Distributed \\deep learning} & \tabincell{c}{Gaussian \\mechanism} & Preserve privacy & \tabincell{c}{Preserve the privacy of \\deep neural networks \\with non-convex \\objectives} & \tabincell{c}{Effective in \\a limited number of \\deep neural networks}\\ \hline
Heikkila et al. \cite{Heikkila17} & \tabincell{c}{Distributed \\deep learning} & \tabincell{c}{Gaussian \\mechanism} & Preserve privacy & \tabincell{c}{Achieve DP in \\distributed settings} & \tabincell{c}{Sacrifice \\learning performance} \\ \hline
Cheng et al. \cite{Cheng18} & \tabincell{c}{Distributed \\deep learning} & \tabincell{c}{Gaussian \\mechanism} & Preserve privacy & \tabincell{c}{Low differential \\privacy budget and \\high learning accuracy} & \tabincell{c}{High communication \\overhead}\\ \hline
Zhao \cite{Zhao18} & \tabincell{c}{Distributed \\deep learning} & \tabincell{c}{Gaussian \\mechanism} & Preserve privacy & \tabincell{c}{Use the teacher-student \\paradigm to improve \\learning performance} & \tabincell{c}{High communication \\overhead}\\ \hline
Jayaraman et al. \cite{Jay18} & \tabincell{c}{Distributed \\deep learning} & \tabincell{c}{Zero-concentrated \\DP mechanism} & Preserve privacy & \tabincell{c}{Both output and gradient \\are protected with \\reduced noise} & \tabincell{c}{Data owners' utility \\cannot be maximized}\\ \hline
Zhao et al. \cite{Zhao19} & \tabincell{c}{Distributed \\deep learning} & \tabincell{c}{Exponential and \\Laplace mechanism} & \tabincell{c}{Preserve privacy and \\improve stability} & \tabincell{c}{Preserve the privacy \\of collective deep \\learning systems with \\the existence of \\unreliable participants} & \tabincell{c}{Accuracy is less than \\the centralized methods}\\ \hline
Phan et al. \cite{Phan19} & \tabincell{c}{Distributed \\deep learning} & \tabincell{c}{Gaussian \\mechanism} & \tabincell{c}{Preserve privacy} & \tabincell{c}{Achieve tight \\robustness bound} & \tabincell{c}{Model accuracy \\is sacrificed}\\ \hline
Geyer et al. \cite{Geyer17} & \tabincell{c}{Federated learning} & \tabincell{c}{Gaussian \\mechanism} & \tabincell{c}{Preserve privacy} & \tabincell{c}{Balance tradeoff \\between privacy loss \\and model performance} & \tabincell{c}{Model performance \\depends on the \\number of clients}\\ \hline
Shi et al. \cite{Shi17} & \tabincell{c}{Federated learning} & \tabincell{c}{Concept of \\differential privacy} & Preserve privacy & \tabincell{c}{No P2P communication \\and fault tolerance} & \tabincell{c}{Focus only on \\multi-input functions}\\ \hline
Agarwal et al. \cite{Agarwal18} & \tabincell{c}{Federated learning} & \tabincell{c}{Gaussian and \\binomial mechanisms} & Preserve privacy & \tabincell{c}{Achieve both \\communication efficiency \\and differential privacy} & \tabincell{c}{The analysis of \\binomial mechanism \\may not be tight} \\ \hline
\end{tabular}}
	\label{tab:summaryDL}
\end{table*}

Table \ref{tab:summaryDL} summarizes the papers that apply differential privacy to distributed deep learning and federated learning.
In this summary, we can see that most of these papers make use of the Gaussian mechanism.
This is because the probability density function of the Gaussian distribution is differentiable,
and this property is necessary for calculating the gradient of a learning model.
The Laplace mechanism does not have this property, so that it was seldom to be applied in deep learning.

It is worth pointing out that simply using differential privacy mechanisms during a learning process may not
provide enough security to protect the privacy of a learning model or the training data.
This is because an adversary, who is pretending to be an honest participant,
can use a GAN \cite{Goodfellow14} to generate prototypical samples
of a victim's private training dataset,
and because the generated samples have the same distribution as the victim's training dataset,
the adversary can bypass the protection of differential privacy \cite{Hitaj17}.
A potential solution against this security issue
is to use local differential privacy.
Unlike regular differential privacy which takes into account
all users' data in a dataset,
local differential privacy add randomization on single user's data \cite{Erlingsson14}.
Thus, local differential privacy has a finer granularity and stronger privacy guarantee.
Even if an adversary has the access to the personal responses of
an individual user in the dataset,
the adversary is still unable to learn accurate information about the user's personal data.
% add the description on how to fix above problem.

% add a further possible development of dp in DL. You can add your two new papers idea here.

Moreover, there are two urgent research problems
which need further investigation.
The first direction is model inversion attack and its defense \cite{Fred15,Yang19ccs}.
Model inversion attacks aim to infer the training data from a target model's predictions.
To implement model inversion attacks,
a popular method is to train a second model called attack model \cite{Yang19ccs}.
The attack model takes the target model's predictions as input
and outputs reconstructed data
which are expected to be very similar to the training data of the target model.
Most of existing defense methods focus only on membership inference attacks
so that their effectiveness on model inversion attacks is still unclear.
A potential defense method against model inversion attacks
is to adopt differential privacy to modify the target model's predictions.
The major reason of the success of model inversion attacks is owning to
the redundant information contained in the target model's predictions.
Thereby, if this redundant information can be destroyed,
the attacks can be effectively defended.

The second direction is the client accuracy in federated learning.
In regular federated learning, the server takes the updates from all clients equally, aiming to minimize an aggregate loss function in general.
However, minimizing an aggregate loss function cannot
guarantee the accuracy of individual clients in the federated network \cite{Mohri19,Li20b},
which is unfair to the clients.
To improve the learning accuracy of the clients, Li et al. \cite{Li20b} introduce an aggregate re-weighted loss function in federated learning,
where different clients are allocated different weights.
A limitation in this type of method is that
the server still need to send all the clients the same model update.
This limitation could be overcome by enabling the server
to send different model updates to different clients
according to each client's requirements.
The server could use joint differential privacy \cite{Kearns14}
to differentiate the model updates in federated learning.

%Fair federated learning aims to achieve a fairly clients updating distribution
%across clients in federated networks \cite{Mohri19,Li20b}.
%a mechanism offers joint differential privacy,
%if for each client $i$, the generation of all the other clients' model updates
%does not affect much on client $i$'s model update.

\vspace{-6mm}

\section{Differential Privacy in Multi-Agent Systems}\label{sec:DP-MAS}
A multi-agent system is a loosely coupled group of agents, such as sensor networks \cite{Ye18},
power systems \cite{Ye11} and cloud computing \cite{Ye19TSC},
interacting with one another to solve complex domain problems \cite{Wooldridge95, Ye17}.
An agent is an intelligent entity that can perceive its environment
and act upon the environment through actuators.
Currently, multi-agent systems face challenges with privacy violation, security issues and communication overhead.

In Mcsherry et al.'s early work~\cite{McSherry07STOC},
differential privacy mechanisms were applied to auctions to diminish the impact of untrusted participants.
In multi-agent systems, differential privacy mechanisms can also avoid malicious agents through a similar mechanism.
There is an increasing trend to apply differential privacy techniques to multi-agent systems
so as to preserve the agents' privacy \cite{Fioretto19} or improve the agents' performance \cite{Pai16}.
This section focuses on some key sub-areas of multi-agent systems, including
multi-agent learning, auction, and game theory.% and data sharing.

\subsection{Differential privacy in multi-agent reinforcement learning}

Multi-agent learning is generally based on the reinforcement learning \cite{Tuyls12}.
Normally, an agent learns proper behavior through the interactions with their environment and other agents in a trial-and-error manner.
Every time an agent performs an action,
they receive a reward which tells them how good that action was for accomplishing the given goal.
Importantly, agents can and do change their strategy to get or better rewards.
Therefore, the aim of the agent is to maximize its long-term expected reward by taking sequential actions.

\begin{figure}[htpb]
\centering
\includegraphics[scale=0.7]{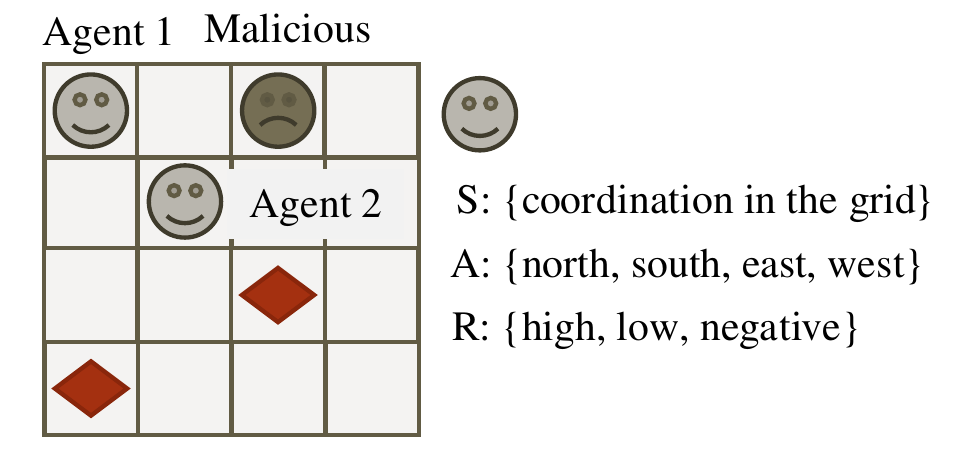}
\caption{A multi-agent learning example}
\label{fig:multiagent}
\end{figure}

For example, Figure~\ref{fig:multiagent} shows a set of sweeper robots (the agents with smiling or crying faces)
who are collecting rubbish from a grid (the red diamonds).
When a robot plans to move to the corner of the grid, it may try to move to the right first.
However, if it bumps into the wall, it receives a very low reward;
thus, the robot learns that moving to the right from its current location is not a good idea.

Standard multi-agent learning approaches may need a large number of interactions between agents and an environment to learn proper behaviors \cite{Bus08}.
Therefore, to improve agent learning performance, agent advising was proposed,
where agents are allowed to ask for advice from each other \cite{Silva18}.
For example, the robot in position $(1,1)$ in Fig. \ref{fig:multiagent} can ask its neighbor in $(2,2)$ for advice
and may obtain the knowledge that it cannot move to the left from its current location.

Existing agent advising approaches, however, suffer from malicious agents
who may provide false advice to hinder the performance of the system,
and heavy communication overheads \cite{Silva17,Silva19}
because agents are allowed to broadcast to all their neighboring agents for advice \cite{Silva17,Silva19}.
%this incurs a large amount of communication overhead.
For example, in Figure~\ref{fig:multiagent}, the malicious robot (the crying face) may provide false information to the other robots
so that the rubbish is not collected in time.

\subsubsection{Differential privacy to improve the security of the reinforcement learning}

Differential privacy mechanisms can provide a security guarantee that
a malicious agent being in or out of a multi-agent system has little impact on the utility of other agents.
As the probability of selecting neighbors to ask for advice is based on the reward history provided by neighbors,
exponential or Laplace mechanisms can be applied to this step to diminish the impact of malicious agents on security purpose.
Moreover, the composition of the privacy budget can naturally control the communication overhead,
namely by limiting the amount of advice allowed throughout the whole system.

Ye et al. \cite{Ye19} proposed a differentially private framework to deal with malicious agents and communication overhead problems.
Using Fig.~\ref{fig:multiagent} as an example,
suppose each agent in the grid environment wants to move to the corner
and has the moving knowledge that can be shared with others.
%\begin{enumerate}
%  \item available actions set $A$ in each location: move to the north, south, east, and west;
%  \item the reward set $R$ of each available action: the reward will be higher if close to the corner, %and will be negative if bumping into a wall;
%  \item a reward vector $S$, which records the rewards gained from using other agents' advice;
%  \item the existence of neighboring agents;
%  \item a communication budget $\epsilon$ to control the communication overhead.
%\end{enumerate}

\begin{figure}[ht]
\centering
	\includegraphics[scale=0.8]{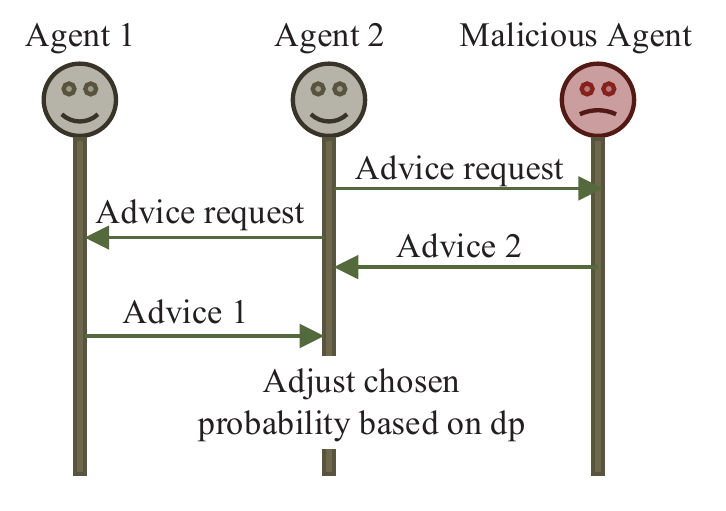}
	\caption{Differentially private multi-agent system}
	\label{fig:advising}
\end{figure}
Fig.~\ref{fig:advising} illustrates the process of agent interaction in this example.
Agent $2$ send out an advice request to its neighbors.
Agent $1$ and the malicious agent would give advice to agent $2$.
Agent $1$'s advice will include the best action in agent $2$'s state according to agent $1$'s knowledge.
However, the malicious agent will always give false advice.
After receiving advice from both neighbors, agent $2$
performs a differentially private adviser selection algorithm, which
applies an exponential mechanism to adjust the chosen probabilities.
Because the malicious agent always provides false advice,
the exponential mechanism may filter its advice with a high probability.
So that the impact of the malicious agent will be diminished.
The system will stop communicating when the privacy budget is used up,
so that the privacy budget can be used to control the communication overhead.

\subsubsection{Summary of differential privacy in reinforcement learning}

Differential privacy technology has been proven
to improve the performance of agent learning in addition to its use as a privacy-preservation tool.
Compared to the benchmark broadcast-based approach,
the differential privacy approach achieves a better performance (normally evaluated by the total reward of the system and the convergence rate) with less communication overhead
when malicious agents are present.
However, further exploration of differential privacy technique in new environments, such as a dynamic or an uncertain environment, would be worthwhile.

\subsection{Differential privacy in auction}

Auction-based approaches are effective in multi-agent systems for task and resource allocation \cite{Parsons11}.
Normally, there are three parties in an auction,
including a seller, an auctioneer and a group of buyers.
The auctioneer presents the item and
receives bids from the buyers following a pre-established bidding convention.
This convention determines the sequence of bids as well as
the way to decide the winner and the price.
%Recently, a number of truthful auction mechanisms have been proposed \cite{Shen19}.
%Truthfulness can motivate bidders to bid their true valuations to avoid market operations.
%However, these true valuations contain bidders' commercial secrets,
%which need to be protected~\cite{Huang15,Alvarez19}.
The current privacy-preserving mechanisms are mainly based on cryptography and multi-party secure computation.
%In these mechanisms, bidders' privacy can be protected during the process of auction computation,
However, there may still be privacy leaks
if one infers their sensitive information from the auction's outcomes.
Differential privacy techniques can help to combat this issue \cite{Chen19}.
The current differentially private auction mechanisms are mostly designed for spectrum allocation in wireless communication.
Radio spectrum is a scarce resource
and thus, the allocation of it needs to be managed carefully.
%As mentioned above, auctions are an efficient and fair way to allocate a resource,
%which is why they are widely used for spectrum allocation.
However, communication also comes with a high desirability for security,
resulting in several private spectrum auction mechanisms.

Zhu et al. \cite{Zhu14} proposed a differentially private spectrum auction mechanism with approximate revenue maximization.
Their mechanism consists of three steps.
First, the auctioneer partitions the bidders into groups and subgroups.
Second, the auctioneer initializes the set of prices
and calculates the probability distribution over these prices using the exponential mechanism.
Finally, based on the probability distribution, the auctioneer randomly selects a price as the auction payment,
and the corresponding bidders are pronounced the winners.

Zhu and Shin \cite{Zhu15} %propose a differentially private spectrum auction mechanism,
alternative achieves strategy-proofness and is polynomially tractable.
The mechanism performs an auction in four steps.
The auctioneer first partitions bidders into groups.
The auctioneer then creates virtual channels for bidders based on their geographical locations and a conflict graph.
Next, the auctioneer computes the probability of selecting each bidder as the winner.
Finally, the auctioneer selects the winner based on an exponential mechanism and determines the payment for the winner.
%\begin{figure}[ht]
%\centering
%    \includegraphics[scale=0.6]{DDSM1.png}
%    \caption{Welfare from basic and improved DDSM}
%\end{figure}
%\begin{figure}[ht]
%\centering
%    \includegraphics[scale=0.6]{DDSM2.png}
%    \caption{Welfare from basic DDSM with different e}
%\end{figure}
% M:200
% N:1050
% G:35
% e1:0.5
% e2:0.5
% max_sell_price:30
% max_buy_price:50
% iter：3000

Wu et al. \cite{Wu16} developed a differentially private auction mechanism
which guarantees bid privacy and achieves approximate revenue maximization.
In their mechanism, the auctioneer first groups bidders based on their conflict graph,
and next determines the price for winners in each group using an exponential mechanism.
Finally, the auctioneer selects the winner based on sorted group revenues.

Chen et al. \cite{Chen19} designed a differentially private double spectrum auction mechanism.
Their mechanism is a uniform price auction mechanism,
where all sellers are paid with a selling clearing price
and all buyers groups are charged with a buying clearing price.
They apply an exponential mechanism twice,
once to select a selling clearance price and again to select a buying clearance price.

In addition to spectrum allocation,
differentially private auction mechanisms have also been developed for resource allocation in cloud computing.
Xu et al. \cite{Xu17} proposed a differentially private auction mechanism for trading cloud resources,
which preserves the privacy of individual bidding information and achieves strategy-proofness.
Their mechanism iteratively progresses through a set of rounds,
where each round consists of four steps.
The auctioneer first uses the exponential mechanism to compute the probability distribution over the set of current bids.
Next, the auctioneer randomly selects a bid from the set as the winner in the current round.
The auctioneer then creates a payment scheme for the winner.
Finally, the winner is removed from the set.

\subsubsection{Summary of differential privacy in auctions}
Although differential privacy in auctions has been widely accepted,
these approaches typically assume a seller or an auctioneer can directly interact with all potential buyers.
This assumption, however, may not be applicable to some real situations,
where sellers and buyers are organized in a network, e.g., social networks \cite{Li17,Zhao18b}.
Auctions in social networks introduce new challenging privacy issues.

The first issue is the bidding propagation.
In a social network, a bid from a buyer cannot be sent directly to the seller
but has to be propagated by other agents to the seller.
These intermediate agents may be potential buyers
and are thus competitive to that buyer.
Therefore, the bid value of that buyer is private and cannot be disclosed to others.
An intuitive way to protect the privacy of that buyer is to add Laplace noise to the bid value.
However, this does not stop the problem of a seller receiving a fake bid and making a wrong decision.
%For example, a buyer $a$ bids $\$ 10$ for a product
%while another buyer $b$ bids $\$ 11$ for the same product.
%However, after adding Laplace noise, buyer $a$'s bid becomes $\$ 11$
%while buyer $b$'s bid becomes $\$ 10.5$.
%The seller then may make a wrong decision to sell the product to buyer $a$.
%Another approach is to adopt a cryptography technique to encrypt the bid for buyers.
%Then, the bid value does not change,
%but managing cipher-keys in a competitive environment becomes an added task.

The second issue is social relationships.
During bid propagation, a trajectory forms
which indicates who is engaged in the propagation.
By investigating this trajectory, the seller may learn things about the buyer's social relationships.

\subsection{Differential privacy in game theory}

Game theory is a mathematical model used to study the strategic interaction among multiple agents or players \cite{Bona18}.
Game theory has been broadly studied and applied in various domains,
e.g., economic science \cite{Acquisti16}, social science \cite{Bona18} and computer science \cite{Durlauf10}.
Most of these studies, however, overlook the privacy of agents, the malicious agents or the stability of the game playing process.
To tackle these issues, differential privacy techniques have been introduced into game theory \cite{Roth13,Pai13,Wanger18}.

Differential privacy-based game theory research can be roughly classified into two categories:
%theoretical research and practical research.
1) using differential privacy to improve the performance of game theory, e.g., stability and equilibrium,
and 2) applying differential privacy to preserve the privacy of agents in games.
%Theoretical research focuses on reveal the facts
%which are caused by the combination of differential privacy and game theory.
%In comparison, practical research focuses on using game theory to model real-world problems
%and developing differential privacy-based mechanisms to address these problems.

%theory, preserving privacy
%The theoretical research focuses on the preserving the privacy,
%improving performance, improving stability and defense the malicious users.
\subsubsection{Differential privacy to improve the performance}
%theory, improving performance and preserving privacy
Kearns et al. \cite{Kearns14} developed a differentially private recommender mechanism for incomplete information games.
Thanks to differential privacy techniques, their mechanism achieves equilibria of complete information games
even with a large number of players and any player's action only slightly affects the payoffs for other players.
Their mechanism offers a proxy that can recommend actions to players.
Players are free to decide whether to opt in to the proxy,
but if players do opt in, they must truthfully report their types.
In addition to satisfy the game-theoretic properties, the mechanism also guarantees player privacy,
namely that no group of players can learn much about the type of any player outside the group.

%theory, improving performance and avoid malicious users
Rogers and Roth \cite{Rogers14} improved the performance and defense of malicious users by expanding on Kearns et al.'s work \cite{Kearns14}
to allow players to falsely report their types
even if the players opt in to the proxy.
They theoretically show that by using differential privacy
to form an approximate Bayes-Nash equilibrium,
players have to truthfully report their types and faithfully follow the recommendations.

%theory, improving performance
Pai et al. \cite{Pai16} improved game performance as well.
They studied infinitely repeating games of imperfect monitoring with a large number of players,
where players observe noisy results, generated by differential privacy mechanisms, about the play in the last round.
The authors find that, theoretically, folk theorem equilibria may not exist in such settings,
which concern all the Nash equilibria of an infinitely repeated game.
Based on this finding, they yield antifolk theorems~\cite{Antifolk},
where restrictions are imposed on the information pattern of repeated games
such that individual deviators cannot be identified \cite{Masso89}.

%theory, improving stability
Lykouris et al. \cite{Lyk16} increased game stability by analyzing the efficiency of repeated games in dynamically changing environments and population sizes.
They draw a strong connection between differential privacy and the high efficiency of learning outcomes in repeated games with frequent change.
Here, differential privacy is used as a tool to find solutions
that are close to optimal and robust to environmental changes.

%application, avoid malicious users
Han et al. \cite{Han15} developed an approximately truthful mechanism to defend against malicious users in an application to manage the charging schedules of electric vehicles.
To ensure users to truthfully report their specifications,
their mechanism takes advantage of joint differential privacy
which can limit the sensitivity of the scheduling process to changes in user specifications.
Therefore, any individual user cannot benefit from misreporting his specification,
which results in truthful reports.

\subsubsection{Applying differential privacy to preserve the privacy}

%The series of application research mainly focus on the privacy preservation and defense of malicious users.

Hsu et al. \cite{Hsu13} modelled the private query-release problem in differential privacy as a two-player zero-sum game between a data player and a query player.
Each element of the data universe for the data player is interpreted as an action.
The data player's mixed strategy is a distribution over her databases.
The query player has two actions for each query.
The two actions are used to penalize the data player,
when the approximate answer to a query is too high or too low.
An offline mechanism, based on the Laplace mechanism, is then developed to achieve the private equilibrium of the game.

%application, preserving privacy
Zhang et al. \cite{Zhang16} developed a general mobile traffic offloading system.
They used the Gale-Shapley algorithm~\cite{Galetheory} to optimize the offloading station allocation plan for mobile phone users.
In this algorithm, to protect users' location privacy, they proposed two differentially private mechanisms based on a binary mechanism \cite{Chan11}.
The first mechanism is able to protect the location of privacy of any individual user
when all the other users are colluding against this user, but the administrator is trusted.
The second mechanism is stronger than the first mechanism
because it assumes that even the administrator is untrusted.

%application, preserving privacy
Zhou et al. \cite{Zhou17} adopted an aggregation game to model spectrum sharing in large-scale and dynamic networks,
where a set of users compete for a set of channels.
They then applied differential privacy techniques to guarantee the truthfulness and privacy of the users.
Specifically, they use a Laplace mechanism to add noise to the cost threshold and users' costs to protect this information.
Moreover, they use an exponential mechanism to decide the mixed strategy aggregative contention probability distribution for each user
so as to preserve the privacy of users' utility functions.

%In addition to preserve agents' privacy, differential privacy is also used in game theory to
%improve agents' performance, increase the stability of games and reduce the impact of malicious agents.

\subsubsection{Summary of differential privacy in game theory}

The current research on differential privacy in game theory has mainly focused on static environments.
These same issues in dynamic environments are generally still open.
%The research on differential privacy in game theory in dynamic environments is still an open issue.
Of the little research that does consider dynamic environments \cite{Lyk16},
only changes in population are considered
while overlooking changes in other areas,
such as the strategies available to each agent
or the utility of each of those strategies.
Studying game theory with changing available strategies is a challenging issue,
as these types of changes may result in no equilibria between agents.
This is because no matter which strategy is taken by an agent,
other agents may always have strategies to defeat that agent.
In other words, other agents are incentivized to unilaterally change their strategies.
However, since differential privacy can be used to force agents to report truthfully,
it may also be used to force agents to reach equilibria.

\subsection{Summary of multi-agent systems}
\begin{table*}[!ht]\scriptsize
\newcommand{\tabincell}[2]{\begin{tabular}{@{}#1@{}}#2\end{tabular}}
	\centering
	\caption{Summary of differential privacy in multi-agent systems}
\scalebox{0.9}{
\begin{tabular}{|c|c|c|c|c|c|} \hline
\textbf{Papers}&\textbf{Research areas}&\textbf{Techniques used}&\textbf{Research aims}&\textbf{Advantages}&\textbf{Disadvantages}\\ \hline
Ye et al. \cite{Ye19} & Multi-agent learning & \tabincell{c}{Laplace \\mechanism} & \tabincell{c}{Avoid \\malicious agents} & \tabincell{c}{Avoid malicious agents \\with low communication \\and computation overhead} & \tabincell{c}{Malicious agents \\cannot be identified}\\ \hline
Zhu et al. \cite{Zhu14} & Auction in spectrum & \tabincell{c}{Exponential \\mechanism} & Preserve privacy & \tabincell{c}{Guarantee both the \\truthfulness of bidders' \\valuations and \\their privacy} & \tabincell{c}{Only approximate \\revenue maximization}\\ \hline
Zhu and Shin \cite{Zhu15} & Auction in spectrum & \tabincell{c}{Exponential \\mechanism} & Preserve privacy & \tabincell{c}{Preserve the privacy of \\ both bidders and \\the auctioneer together} & \tabincell{c}{Only near optimal \\revenue achieved} \\ \hline
Wu et al. \cite{Wu16} & Auction & \tabincell{c}{Exponential \\mechanism} & Preserve privacy & \tabincell{c}{Guarantee both \\bid privacy and fairness} & \tabincell{c}{Only approximate \\revenue
maximization}\\ \hline
Chen et al. \cite{Chen19} & Auction in spectrum & \tabincell{c}{Exponential \\mechanism} & Preserve privacy & \tabincell{c}{Preserve the privacy \\of bidders in \\double spectrum auctions} & \tabincell{c}{Only approximate social \\welfare maximization}\\ \hline
Xu et al. \cite{Xu17} & \tabincell{c}{Auction in \\cloud computing} & \tabincell{c}{Exponential \\mechanism} & Preserve privacy & \tabincell{c}{Preserve the privacy of \\consumers in \\cloud environments} & \tabincell{c}{Only approximate \\truthfulness and revenue \\maximization guarantees}\\ \hline
Hsu et al. \cite{Hsu13} & \tabincell{c}{Game theory in \\databases} & \tabincell{c}{Laplace \\mechanism} & Preserve privacy & \tabincell{c}{Preserve the privacy of \\both individuals and \\analysts of \\database systems} & \tabincell{c}{Achieve only nearly \\optimal error rates}\\ \hline
Kearns et al. \cite{Kearns14} & Game theory & \tabincell{c}{Concept of \\differential privacy} & \tabincell{c}{Preserve privacy and \\Improve performance} & \tabincell{c}{Implement equilibria of \\complete information \\games in settings of \\incomplete information} & \tabincell{c}{The type of a player \\is still possible \\to be revealed}\\ \hline
Rogers and Roth \cite{Rogers14} & Game theory & \tabincell{c}{Concept of \\differential privacy} & \tabincell{c}{Preserve privacy and \\avoid malicious agents} & \tabincell{c}{Implement equilibria of \\complete information \\games in settings of \\incomplete information \\even if players \\are lying} & \tabincell{c}{The type of a player \\is still possible \\to be revealed}\\ \hline
Zhang et al. \cite{Zhang16} & \tabincell{c}{Game theory in \\mobile communication} & \tabincell{c}{Binary \\mechanism} & Preserve privacy & \tabincell{c}{Preserve each user's \\location privacy even if \\other users collude} & \tabincell{c}{The system administrator \\is required to be \\honest or semi-honest}\\ \hline
Zhou et al. \cite{Zhou17} & \tabincell{c}{Game theory in \\spectrum sharing} & \tabincell{c}{Laplace and \\exponential \\mechanisms} & Preserve privacy & \tabincell{c}{Guarantee both \\truthfulness and privacy \\of users} & \tabincell{c}{Achieve only \\approximate Nash \\equilibrium} \\ \hline
Pai et al. \cite{Pai16} & Game theory & \tabincell{c}{Concept of \\differential privacy} & Improve performance & \tabincell{c}{Quantify limit results \\for repeated games} & \tabincell{c}{Achieve only \\approximate equilibria}\\ \hline
Lykouris et al. \cite{Lyk16} & Game theory & \tabincell{c}{Concept of \\differential privacy} & Improve stability & \tabincell{c}{Connect differential \\privacy with learning \\efficiency in \\dynamic games} & \tabincell{c}{The solution is \\approximate optimal}\\ \hline
Han et al. \cite{Han15} & \tabincell{c}{Game theory in \\electric vehicles} & \tabincell{c}{Laplace \\mechanism} & Avoid malicious agents & \tabincell{c}{Reduce the incentive \\of user misreporting} & \tabincell{c}{Achieve only \\approximate truthfulness} \\ \hline
\end{tabular}}
	\label{tab:summaryMAS}
\end{table*}

Table \ref{tab:summaryMAS} summarizes the papers that apply differential privacy to multi-agent systems.
In this summary, three important facts are involved.
First, some of these papers use differential privacy not to preserve the privacy of agents
but for other aims, e.g., avoiding malicious agents and improving agents' performance.
This implies that the differential privacy technique is able to achieve other research aims besides privacy preservation.
In keeping with this spirit, more potential applications of differential privacy are worthy of research.
Second, by using differential privacy, the common disadvantage is that
only approximate optimal results can be achieved.
Thus, more efficient differential privacy mechanisms need to be developed.

Third, most of these papers involve agent interaction; and differential privacy is adopted to guarantee the privacy of interaction information.
Therefore, other multi-agent research, which involves agent interaction,
may also enjoy the benefits of differential privacy
and deserves further investigation.
For example, multi-agent negotiation enables multiple agents to alternatively provide offers
to reach agreements on given events or goods \cite{Dimo19}.
However, offers may explicitly or implicitly contain agents' sensitive information,
e.g., commercial secrets, which should be protected.
Another example is multi-agent resource allocation.
To allocate resources fairly, agents have to reveal their preference
over different types of resources to others \cite{Beynier19}.
The preference, however, might be what the agents incline to hide.
In summary, differential privacy has a great potential to solve diverse problems in multi-agent system.

% add a future development of DP in MAS

\section{Future Research Directions}
%Differential privacy techniques have been widely adopted in various areas of AI to achieve privacy preservation, guarantee security, improve stability, maintain fairness, and ensure truthful reporting. However, some sub-areas, which can also benefit from using differential privacy, have not been thoroughly investigated yet.
%Therefore, we summarize the point of using differential privacy in these sub-areas.

%\begin{table}[!ht]\scriptsize
%\newcommand{\tabincell}[2]{\begin{tabular}{@{}#1@{}}#2\end{tabular}}
%	\centering
%	\caption{Summary of the future directions}
%\begin{tabular}{|c|c|c|} \hline
%\textbf{Research areas}&\textbf{Research aims}&\textbf{Potential %techniques}\\ \hline
%\tabincell{c}{Deep reinforcement \\learning} & \tabincell{c}{Improve %sampling \\efficiency} & \tabincell{c}{Small database \\mechanism} \\ %\hline
%Meta-learning & \tabincell{c}{Preserve privacy} & \tabincell{c}{Gaussian %\\mechanism} \\ \hline
%\tabincell{c}{Generative adversarial \\networks} & \tabincell{c}{Preserve privacy} & \tabincell{c}{Gaussian \\mechanism} \\ \hline
%\tabincell{c}{Multi-agent \\advising learning} & \tabincell{c}{Improve flexibility \\of advising} & \tabincell{c}{Laplace \\mechanism}\\ \hline
%\tabincell{c}{Multi-agent \\transfer learning} & \tabincell{c}{Preserve privacy} & \tabincell{c}{Laplace and \\exponential mechanisms}\\ \hline
%\tabincell{c}{Multi-agent \\planning} & \tabincell{c}{Preserve privacy} & \tabincell{c}{Laplace \\mechanism} \\ \hline
%\end{tabular}
%	\label{tab:futurework}
%\end{table}

%\subsection{Machine learning}
\subsection{Private transfer learning}

In addition to introducing differential privacy into standalone machine learning,
differentially private transfer learning has also been investigated \cite{Xie17,Wang18}.
Transfer learning aims to transfer knowledge from source domains to improve learning performance in target domains \cite{Wang18}.
It is typically used to handle the situation that
data are not stored in one place but distributed over a set of collaborative data centers \cite{LeTien19,Yao19}.
For example, transfer learning can be used in speech recognition
to transfer the knowledge of connectionist temporal classification model
to the target attention-based model to overcome the problem of limited speech resource in the target domain \cite{Qin18}.
Transfer learning can also be used in recommendation systems
to address the data-sparsity issue by enabling knowledge to
be transferred among recommendation systems \cite{Zhao13}.
Instead of transferring raw data, the intermediate computation results are transferred
from source domains to target domains.
However, even the intermediate results are potentially vulnerable to privacy-breach \cite{Wang09},
which is the motivation of privacy-preserving transfer learning.

\subsection{Deep reinforcement learning}
Deep reinforcement learning is a combination of reinforcement learning and deep learning \cite{Francois18},
and could be used to solve a wide range of complex decision-making problems that were previously beyond the capability of regular reinforcement learning. The learning process of deep reinforcement learning is similar to regular reinforcement learning in that both are based on trial-and-error. However, unlike regular reinforcement learning which may use a reward value table (Q-table) to store learned knowledge, deep reinforcement learning uses a deep Q-network instead. One of the advantages of using a deep Q-network is that deep reinforcement learning can take high-dimensional and continuous states as inputs, which is close to unfeasible with regular reinforcement learning.

Differential privacy in deep reinforcement learning has not been researched thoroughly. Wang and Hegde~\cite{Wang19} applied differential privacy to deep reinforcement learning to protect the value function approximator by adding Gaussian to the objective function, but their work still focuses on the “deep learning” aspects of the approach rather than the “reinforcement learning” parts. Compared to standard reinforcement learning and deep learning, deep reinforcement learning has some unique features. First, the training samples are collected during learning rather than pre-assembled before learning. Second, the training samples may not be independent but rather highly correlated. Third, the training samples are not usually labelled. Thus, to avoid overfitting with deep reinforcement learning, experience replay is required, which means randomly selecting a set of samples for training in each iteration. As discussed in the previous sections, differential privacy can improve the stability of learning. Therefore, it may be interesting to research whether introducing differential privacy into deep reinforcement learning can help to avoid overfitting.

\subsection{Meta-learning}

Meta-learning, also known as ‘learning to learn’, is a learning methodology that systematically observes how different machine learning approaches perform on a wide range of learning tasks and then learning from these observations~\cite{Vilalta02, Lemke15,Van19}. In meta-learning, the goal of the trained model is to quickly learn a new task from a small amount of new data. Also, the trained model should be able to learn on a number of different tasks~\cite{Finn17,Nichol18}, but this opens the risk of breaching the privacy of the different task owners~\cite{Li20}.

Recently, Li et al.~\cite{Li20} introduced differential privacy into meta-learning to preserve the privacy of task owners. Specifically, they use a certified $(\epsilon,\delta)$-differential privacy stochastic gradient descent~\cite{Bassily19} with each task, which guarantees that the contribution of each task owner carries global differential privacy guarantees with respect to the meta-learner. However, to guarantee global differential privacy, the number of tasks has to be known beforehand. This is hard to know in some situations, such as online meta-learning where tasks are revealed one after the other in a dynamic manner~\cite{Finn19}. Therefore, it would be worthwhile developing a new differential privacy-based algorithm to preserve the privacy of task owners in online meta-learning,

\subsection{Generative adversarial networks}
Generative adversarial networks (GANs) \cite{Goodfellow14} are a framework for producing a generative model by way of a two-player minimax game.
One player is the generator
who attempts to generate realistic data samples by transforming noisy samples
drawn from a distribution
using a transformation function with learned weights.
The other player is the discriminator
who attempts to distinguish between synthetic data samples created by the generator.
%and the real data samples drawn from an actual dataset using a function with learned weights.

The GAN framework is one of the most successful learning models and has been applied to applications such as imitating expert policies \cite{Ho16} and domain transfer \cite{Yoo16}. More recently, GANs have been extended to accommodate multiple generators and discriminators so as to address more complex problems.
%For example, Ghosh et al. %\cite{Ghosh18} developed a multi-agent GAN architecture that includes multiple generators and one discriminator.
%The generators are used to capture diverse, high-probability classes, while the discriminator is designed to not only discern between real and fake samples but also to identify the generator producing the fake samples. In this way, the multi-agent GAN architecture is able to disentangle different modalities when trained using a highly-challenging diverse-class dataset.
%Another example is Durugkar et al. \cite{Duru17}, who proposed a generative multi-adversarial network (GMAN), which involves multiple discriminators. These discriminators vary with a range of roles, from being a formidable adversary to being a forgiving teacher. The formidable adversary can well approximate the objective function, and the forgiving teacher can well match the generator’s capabilities. As a result, the GMAN framework learns faster than the standard GAN model. Further, in image generation tasks, GMAN frameworks can produce higher quality samples than the standard GAN model.
Like other learning models, GAN frameworks also suffer from the risk of information leaks. More specifically, the generator model estimates the underlying distribution of a dataset and randomly generates realistic samples, which means the generator, through the power of deep neural networks, remembers training samples. Now, when the GAN model is applied to a private or sensitive dataset, the privacy of the dataset may be leaked. To deal with this problem, Xu et al. proposed a GAN-obfuscator \cite{Xu19}, i.e., a differentially private GAN framework, where carefully designed Gaussian noise is added to the gradients of learning models during the learning procedure. By using the GAN-obfuscator, an unlimited amount of synthetic data can be generated for arbitrary tasks without disclosing the privacy of training data. However, although the framework can guarantee the privacy of training data, there is only one generator and one discriminator in this framework. Therefore, a useful direction of future research might be to extend these principles to multiple generators and discriminators to address more complex problems.

\subsection{Multi-agent systems}
\subsubsection{Multi-agent advising learning}
When an agent is in an unfamiliar state during a multi-agent learning process, it may ask for advice from another agent \cite{Silva17}. These two agents then form a teacher-student relationship. The teacher agent offers advice to the student agent about which action should be taken. Existing research is based on a common assumption that the teacher agent can offer advice only if it has visited the same state as the student agent’s current state. But this assumption might be relaxed by using differential privacy technique.

The property of differential privacy can be borrowed to address the advice problem. Two similar states are interpreted as two neighbouring datasets. The advice generated from the states is interpreted as the query result yielded from datasets. Since two results from neighbouring datasets can be considered approximately identical, two pieces of advice generated from two similar states can also be considered approximately identical. This property can thus guarantee that advice created in a state can still be used in another similar state. Hence, this may be an interesting way to improve agent learning performance.

\subsubsection{Multi-agent transfer learning}
When agents transfer knowledge between each other to improve learning performance,
a key problem discussed is that privacy needs to be preserved \cite{Silva19}. Existing methods are typically based on homomorphic cryptosystems \cite{Sakuma08,Wu18,Liu19}. However, homomorphic cryptosystems have a high computation overhead and, therefore, may not be very efficient in resource-constrained systems, e.g., wireless sensor networks. Differential privacy, with its light computation overhead, therefore, could be a good alternative in these situations.

%\subsubsection{Multi-agent planning}

%Multi-agent planning is a fundamental research problem in multi-agent systems \cite{desJardins99,Ye17} and has been broadly examined in real-world applications \cite{Ore18,Mah19}. Research on multi-agent planning aims to improve the working efficiency of agents by making plans in advance. During jointly automated planning, agents have to share information with each other, but sharing means risking privacy leaks and, hence, privacy preservation is a factor in multi-agent planning \cite{Shani18}.

%Most existing planning approaches are either weak on privacy preservation, or they entirely overlook privacy issues \cite{Maliah17,Torreno17}. Very few approaches offer strong privacy guarantees \cite{Brafman15}, but even the ones that do may not provide strong privacy as well as completeness and efficiency as summarized in \cite{Tozicka17}. Introducing differential privacy into multi-agent planning systems could, however, provide all three.

\subsubsection{Multi-agent reasoning}
Reasoning is an ability that enables an agent to use known facts to deduce new knowledge.
It has been widely employed to address various real-world problems.
For example, knowledge graph-based reasoning can be used in speech recognition to parse speech contents into logical propositions \cite{Zhou20},
and case-based reasoning can be adopted to address the data-sparsity issue in recommendation systems
by filling in the vacant ratings of the user-item matrix \cite{Tawfik17}.
A typical reasoning method is based on the Belief, Desire and Intention (BDI) model \cite{Inverno04}.
An agent's beliefs correspond to information the agent has about the world.
%The representation of an agent's beliefs is usually symbolic, such as propositions.
%An agent's desires are its goals, i.e. the tasks that the agent has to complete.
%Since an agent may not achieve all its desires,
%it has to choose some subset of available desires.
%These chosen desires are intentions.
%To operationalize the intention model, an agent is implemented a plan library,
%which is a set of plans that may be undertaken by the agent to achieve its intentions.

Reasoning is a powerful tool in AI especially
when it is combined with deep neural networks.
For example, Mao et al. \cite{Mao19} has recently proposed a neuro-symbolic concept learner
which combines symbolic reasoning with deep learning.
Their model can learn visual concepts, words and semantic parsing of sentences
without explicit supervision on any of them.
As reasoning requires querying known facts
which may contain private information,
privacy preservation becomes an issue in reasoning process.
Tao et al. \cite{Tao14} propose a privacy-preserving reasoning framework.
Their idea is to hide the truthful answer from a querying agent
by providing the answer ``Unknown'' to a query.
Then, the querying agent cannot distinguish between the case
that the query is being protected and the case that
the query cannot be inferred from the known facts.
However, simply hiding truthful answers may seriously hinder the utility of querying results.
Differential privacy, with its theoretical guarantee of utility of querying results,
may be a promising technique for privacy-preserving reasoning.

Recently, there have been great efforts to combine differential privacy with reasoning \cite{Barthe13,Barthe16,Zhang19}.
These works, however, aim to take advantage of reasoning to prove
differential privacy guarantees of programs,
instead of using differential privacy to guarantee privacy-preserving reasoning.
Therefore, a potential direction of future research may be introducing
differential privacy into reasoning process to guarantee
the privacy of known facts.

%This is because differential privacy has a strong theoretical foundation
%which can guarantee the privacy of information without sacrificing the utility of information.
%This property of differential privacy perfectly matches the requirements of privacy-preserving multi-agent planning.

\subsection{Combination of machine learning, deep learning and multi-agent systems}
A novel research area by combining
machine learning, deep learning and multi-agent systems
is the multi-agent deep reinforcement learning (MADRL) \cite{Leal19}.
In MADRL, multi-agent system technique is used to coordinate the behaviors of agents;
machine learning technique is responsible for guiding the learning process of agents;
and deep learning is employed by agents to learn efficient strategies.

One of the current research directions along MADRL is the action advising \cite{Ilhan19,Omidshafiei19,Silva20}.
%Unlike the regular multi-agent reinforcement learning, in MADRL,
%the number of states may be very large or even infinite.
%Thus, it is infeasible to use the number of visits to a state
%to measure the confidence in that state
%as usual.
%An efficient way is to adopt a random network distillation
%to measure the confidence in a state
%by measuring the mean squared error between two neural networks:
%the target network and the predictor network \cite{Ilhan19}.}
Action advising in regular multi-agent reinforcement learning allows a teacher agent
to offer only an action as advice to a student agent in a concerned state.
By comparison, action advising in MADRL usually allows a student agent
to query a teacher agent's knowledge base to receive action suggestions \cite{Silva20}. %: $\pi_{\Delta}:S\times A\rightarrow (0,1)$  \cite{Silva20}.
However, as the number of states in MADRL is very large,
an agent's knowledge base may contain the agent's
very rich private information that should be protected.
Privacy-preservation in MADRL is still an open research problem
which may be addressed by using differential privacy.

\section{Conclusion}
In this paper, we investigated the use of differential privacy in selected areas of AI. We described the critical issues facing AI and the basic concepts of differential privacy, highlighting how differential privacy can be applied to solving some of these problems. We discussed the strengths and limitations of the current studies in each of these areas and also pointed out the potential research areas of AI where the benefits of differential privacy remain untapped. In addition to the three areas of focus in this article – machine learning, deep learning and multi-agent learning – there are many other interesting areas of research in AI that have also leveraged differential privacy, such as natural language processing, computer vision, robotics, etc. Surveying differential privacy in these areas is something we intend to do in future work.

%\ifCLASSOPTIONcompsoc
%\else
%\section*{Acknowledgment}
%\fi
%This work was supported by the National
%Natural Science Foundation of China (No. 61972366),
%and in part by NSF under grants III-1526499, III-1763325, III1909323, and CNS-1930941.

\bibliographystyle{IEEEtran}
{\small \bibliography{references}}

\ifCLASSOPTIONcaptionsoff
  \newpage
\fi

\end{document}